\documentclass[english,american,a4paper,american,american,american,american,a4paper,american,prb,eqsecnum, showpacs,manuscript]{revtex4}
\usepackage[T1]{fontenc}
\usepackage[latin1]{inputenc}
\usepackage{amsmath}
\usepackage{graphicx}
\usepackage{amssymb}

\makeatletter

\makeatletter
\newcommand{\Nomega}{N}
\newcommand{\Vplus}{{V}_{+}}
\newcommand{\Vminus}{{V}_{-}}
\newcommand{\Vplusminus}{{V}_{\pm}}


\makeatother

\makeatother

\makeatother

\makeatother

\makeatother

\makeatother

\makeatother

\makeatother

\usepackage{babel}
\makeatother
\begin{document}

\title{Noise spectrum of a tunnel junction coupled to a nanomechanical oscillator}

\author{J. Wabnig}

\affiliation{Department of Physics, Ume\aa{} University, SE-901 87 Ume\aa{}}

\author{J. Rammer }

\affiliation{Department of Physics, Ume\aa{} University, SE-901 87 Ume\aa{}}

\author{A. L. Shelankov}

\affiliation{Department of Physics, Ume\aa{} University, SE-901 87 Ume\aa{}}

\affiliation{A. F. Ioffe Physico-Technical Institute, 194021 St. Petersburg, Russia}

\begin{abstract}
A nanomechanical resonator coupled to a tunnel junction is studied.
The oscillator modulates the transmission of the junction, changing
the current and the noise spectrum. The influence of the oscillator
on the noise spectrum of the junction is investigated, and the noise
spectrum is obtained for arbitrary frequencies, temperatures and bias
voltages. We find that the noise spectrum consists of a noise floor
and a peaked structure with peaks at zero frequency, the oscillator
frequency and twice the oscillator frequency. The influence of the
oscillator vanishes if the bias voltage of the junction is lower than
the oscillator frequency. We demonstrate that the peak at the oscillator
frequency can be used to determine the oscillator occupation number,
showing that the current noise in the junction functions as a thermometer
for the oscillator. 
\end{abstract}


\maketitle

\section{Introduction \label{intro}}

The recent years have seen the shrinking of mechanical components
to micrometer and further to nanometer size, spawning the new field
of nanomechanics. \cite{SchwRou05,Ble04} Manufacturing techniques
borrowed from semiconductor chip manufacturing (see e.g. Ref.~\onlinecite{EkiRou05})
as well as bottom up approaches utilizing nanotubes \cite{Saz04}
make it possible to produce nanomechanical resonators with resonance
frequencies presently demonstrated up to 1 GHz. Nanomechanical resonators
are of interest for several reasons. In proposed tests of the limits
of quantum mechanics one would like to investigate the decoherence
behavior of superpositions of macroscopically distinct states, of
e.g. a nanomechanical resonator.\cite{Leg02,MarSimPen03} Typical
nanomechanical resonators contain a macroscopic number of atoms, $\sim10^{7}$,
making the amplitude of the basic mode a macroscopic observable. The
signature of such superpositions is strongest for low oscillator occupation
numbers. Presently obtainable resonator frequencies are high enough
to make cooling to the ground state feasible. Experimental efforts
are under way to reach the first goal necessary for these measurements,
cooling a nanomechanical resonator to the ground state. \cite{NaiBuuLah06,LahBuuCam04,GigBohZei06,ArcCohBri06,KleBou06}

Other applications include the use of nanomechanical resonators as
ultra-sensitive force detectors. Mass detection with zeptogram resolution
utilizing nanomechanical resonators has been realized only recently.\cite{YanCalFen06}
Nanosized cantilevers can also be used to detect magnetic forces.
Detection of a single electron spin using a nanomechanical cantilever
has already been demonstrated. \cite{RugBudMam04} Nanomechanical
resonators can also find an application in the context of quantum
computing. Coherent mechanical oscillators are suggested as coupling
elements between phase qubits in a solid state quantum computer. \cite{GelCle05}
All the mentioned applications not only require the fabrication of
a suitable oscillator but also a way to detect the motion of a nanomechanical
resonator. Different schemes for detection have been proposed (see
e.g. Ref.~\onlinecite{SchwRou05}). The most promising candidates
for sensitive readout are electrical devices, such as tunnel junctions
or single electron transistors, incorporated on the same chip as the
nanomechanical resonator.\cite{SchwRou05}

In the light of the possible applications it is necessary to obtain
a theoretical understanding of nanomechanical resonators interacting
with electrical devices on a chip. Theoretical descriptions of charge
dynamics influenced by an oscillator have, up to now, mainly used
a master equation technique. Mozyrsky and Martin investigated the
model, where the transmission coefficient of a tunnel junction depends
on the position of a nearby harmonic oscillator, in the zero temperature
limit. They found that the oscillator acquires an effective temperature
proportional to the junction bias voltage and also find the influence
of the oscillator on the junction current at zero temperature.\cite{MozMar02}
Clerk and Girvin calculated the noise induced by a harmonic oscillator
in a tunnel junction for dc and ac bias at zero temperature using
a Markovian master equation.\cite{CleGir04} The authors, together
with Khomitsky found the current in a tunnel junction influenced by
an oscillator for arbitrary system parameters.\cite{WabKhoRam05}
The current and the noise power spectrum for an asymmetric junction
in the high voltage limit were also calculated. Smirnov, Mourokh and
Horing\cite{SmiMouHor03} analyzed the position fluctuations in the
stationary state of a nanomechanical oscillator coupled to a tunnel
junction for an exponential dependence of the tunneling amplitude
on the oscillator position as well as the current through the tunnel
junction. The theoretical descriptions of a nanomechanical oscillator
interacting with a single electron transistor concentrated on the
effects the single electron transistor, acting as a non-equilibrium
environment, has on the oscillator. Rodrigues and Armour\cite{RodArm05}
derived a master equation for an oscillator-single electron transistor
system and investigated stationary state properties and dynamics.
Blencowe, Imbers and Armour\cite{BleImbArm05} as well as Clerk and
Bennett\cite{CleBen05} considered the interaction of a superconducting
single electron transistor with a nanomechanical resonator and discovered
that for a particular source-drain voltage the single electron transistor
can cool the oscillator and investigated a regime where the damping
constant becomes negative .

In this article we consider a tunnel junction coupled to a harmonic
oscillator as a model for a nanomechanical resonator interacting with
a detector and concentrate on the features of the noise power spectrum
that the oscillator induces in the tunnel junction. We are interested
in this model system for several reasons: The induced noise provides
a means to determine the temperature of the oscillator. A shot noise
thermometer, utilizing a tunnel junction, was demonstrated by L. Spietz
\emph{et al.}\cite{SpiLehSid03} and was shown to work over a temperature
range from $50\,\textnormal{mK}$ to $25\,\textnormal{K}$. The noise
induced by a nanomechanical oscillator in a SET has been successfully
used to detect the temperature of a nanomechanical oscillator in two
recent experiments,\cite{LahBuuCam04,NaiBuuLah06} in the region where
the occupation number of the oscillator was large.

We also want to investigate if there exists a signature of the oscillator
in the noise power spectrum even if the oscillator is in its ground
state and the voltage is insufficient to excite the oscillator. A
recent treatment of a similar system, considering a spin instead of
an oscillator, claims a non-vanishing contribution to the noise power
under these conditions.\cite{LiCuiYan05} A similar prediction could
be made by blithely extending the result obtained by a Markovian master
equation (e.g. from Ref.~\onlinecite{WabKhoRam05}) into the region
where the bias voltage is smaller than the oscillator frequency. Revisiting
the junction/oscillator system utilizing a different technique will
give us opportunity to investigate the question of the seemingly non-vanishing
noise.

Also there has been a recent theoretical discussion about which current-current
correlator is detected in a noise experiment. Lesovik and Loosen,\cite{LesLoo97}
Aguado and Kouwenhoven,\cite{AguKou00} as well as Gavish \emph{et
al.\cite{GavLevImr00}} argue, that a passive detector, e.g. a LC
oscillator at zero temperature or a two-level system, can only detect
the positive frequency part of the Fourier transform of the unsymmetrized
current-current correlator. The oscillator coupled to a tunnel junction
gives us the opportunity to revisit this question in the context of
a more complicated system than a mere tunnel junction.

In this paper we will apply a Green's function technique to calculate
the noise power spectrum of a tunnel junction coupled to an oscillator
in the approximation of weak coupling, but for otherwise arbitrary
parameters. The article is structured as follows: In section \ref{definitions}
we introduce the model Hamiltonian, and in section \ref{stationarystate}
we consider the stationary state of the oscillator using a Green's
function technique. In section \ref{Noise} we calculate the average
current through the junction as well as the unsymmetrized noise power
spectrum. We consider application of the results in section \ref{thermometry},
discussing noise thermometry. We present the conclusions in section
\ref{conclusions}. Details of the calculations are presented in appendices.

\section{Oscillator interacting with a tunnel junction \label{definitions}}

Let us consider the situation of a nanomechanical resonator, modelled
as harmonic oscillator, interacting with a measuring device, modelled
as a tunnel junction. The oscillator modulates the transmission amplitude
of the junction thus changing the current and noise characteristics
of the junction. The biased junction in turn acts as a non-equilibrium
environment for the oscillator, driving the oscillator from its initial
state into a stationary thermal equilibrium state, albeit with a temperature
different from the environment temperature of the tunnel junction.
The Hamiltonian of the model system is \begin{equation}
\hat{H}=\hat{H}_{0}+H_{l}+H_{r}+\hat{H}_{T}\label{ham}\end{equation}
 where $\hat{H}_{0}$ is the Hamiltonian for the isolated harmonic
oscillator with bare frequency $\Omega_{B}$ and mass $m$. A hat
marks operators acting on the oscillator degree of freedom. The Hamiltonians
$H_{l,r}$ specify the isolated left and right electrodes of the junction
\begin{equation}
H_{l}=\sum\limits _{\mathbf{l}}\varepsilon_{\mathbf{l}}\, c_{\mathbf{l}}^{\dagger}c_{\mathbf{l}}\quad,\quad H_{r}=\sum\limits _{\mathbf{r}}\varepsilon_{\mathbf{r}}\, c_{\mathbf{r}}^{\dagger}c_{\mathbf{r}}\label{hamlr}\end{equation}
 where $\mathbf{l,}\mathbf{r}$ label the quantum numbers of the single
particle energy eigenstates in the left and right electrodes, respectively,
with corresponding energies $\varepsilon_{\mathbf{l},\mathbf{r}}$
and annihilation and creation operators. The operator $\hat{H}_{T}$
describes the tunnelling, \begin{equation}
\hat{H}_{T}=\hat{\mathcal{T}}+\hat{\mathcal{T}}^{\dagger}\quad,\quad\hat{\mathcal{T}}=\sum\limits _{\mathbf{l},\mathbf{r}}\hat{T}_{\mathbf{lr}}c_{\mathbf{l}}^{\dagger}c_{\mathbf{r}}\label{hamtum1}\end{equation}
 with the tunneling amplitudes, $\hat{T}_{\mathbf{lr}}=\hat{T}_{\mathbf{rl}}^{\dagger}$,
depending on the oscillator degree of freedom. Due to the interaction
of the tunnel junction and the oscillator, the tunnelling amplitudes
and thereby the conductance of the tunnel junction depend on the state
of the oscillator. In the following we assume linear coupling between
the oscillator position and the tunnel junction \begin{equation}
\hat{T}_{\mathbf{lr}}=v_{\mathbf{lr}}+w_{\mathbf{lr}}\hat{x}\label{tlin}\end{equation}
 where $v_{\mathbf{lr}}=v_{\mathbf{rl}}^{*}$ is the unperturbed tunneling
amplitude and $w_{\mathbf{lr}}=w_{\mathbf{rl}}^{*}$ its derivative
with respect to the position of the oscillator.

To discuss the current and noise in the tunnel junction, the current
operator is needed \begin{equation}
\hat{I}=i\left(\hat{\mathcal{T}}-\hat{\mathcal{T}}^{\dagger}\right).\label{currentop}\end{equation}

The tunneling Hamiltonian consists of a part independent of the state
of the oscillator and a part that depends on the state of the oscillator,
so that the tunneling Hamiltonian, Eq.~(\ref{hamtum1}), can be presented
on the form \begin{equation}
\hat{H}_{T}=h_{v}+\hat{x}h_{w}.\label{tunneling}\end{equation}
 For notational convenience, we have introduced the symbolic notation
\begin{equation}
h_{u}=\mathcal{T}_{u}+\mathcal{T}_{u}^{\dagger},\quad\mathcal{T}_{u}=\sum_{\mathbf{lr}}u_{\mathbf{lr}}c_{\mathbf{l}}^{\dagger}c_{\mathbf{r}},\quad u=v,w\,,\label{hu}\end{equation}
 where the symbol $u_{\mathbf{lr}}$ can take the values $v_{\mathbf{lr}}$
or $w_{\mathbf{lr}}$. Similarly we can write the current operator,
Eq.~(\ref{currentop}), as \begin{equation}
\hat{I}=j_{v}+\hat{x}j_{w},\label{current}\end{equation}
 with

\begin{equation}
j_{u}=i\left(\mathcal{T}_{u}-\mathcal{T}_{u}^{\dagger}\right),\quad u=v,w\,.\label{ju}\end{equation}

When calculating current and noise in the tunnel junction the following
combinations of the model parameters $v_{\mathbf{lr}}$ and $w_{\mathbf{lr}}$
appear \begin{equation}
\left\{ \begin{array}{c}
G_{vv}\\
G_{ww}\\
G_{vw}\end{array}\right\} =2\pi\sum\limits _{\mathbf{lr}}\left\{ \begin{array}{c}
v_{\mathbf{lr}}^{2}\\
w_{\mathbf{lr}}^{2}\\
v_{\mathbf{lr}}w_{\mathbf{lr}}\end{array}\right\} \left(-\frac{\partial f(\varepsilon_{\mathbf{l}})}{\partial\varepsilon_{\mathbf{l}}}\right)\delta(\varepsilon_{\mathbf{l}}-\varepsilon_{\mathbf{r}}),\label{cond}\end{equation}
 where $f$ is the Fermi function. Here and in the following, the
transmission matrix elements are assumed real.

\section{Stationary state properties of the oscillator \label{stationarystate}}

\label{Stationary-state}

In this section we consider the properties of the stationary state
a harmonic oscillator reaches due to interaction with a tunnel junction.
In the Keldysh technique (for a review see e.g. Ref.~\onlinecite{RamSmi86}),
we introduce the contour ordered oscillator matrix Green's function
\begin{equation}
D(\tau,\tau')=-i\left\langle \mathrm{T}_{c}\left(\hat{x}_{H}(\tau)\hat{x}_{H}(\tau')\right)\right\rangle .\label{D}\end{equation}
 The subscript $H$ refers to an operator in the Heisenberg picture.
Each of the times $\tau$ and $\tau'$ belong to one of the two branches
of the Keldysh contour from $-\infty$ to $+\infty$, $\textnormal{T}_{c}$
denotes the contour ordering operator that orders operators along
the Keldysh contour. We will use the symbol $\tau$ to denote times
on the contour, whereas $t$ denotes real times. The branch index
makes $D$ a $2\times2$ matrix. The Keldysh matrix defined by Eq.~(\ref{D})
can be linearly transformed to the {}``triangular'' form: \textbf{\begin{equation}
D=\left(\begin{array}{cc}
D^{R} & D^{K}\\
0 & D^{A}\end{array}\right),\label{jwe}\end{equation}
}  where $D^{R}$, $D^{A}$, and $D^{K}$ are the retarded, advanced
and Keldysh Green's functions, respectively.

For a stationary state, the elements of the Keldysh matrix are functions
of only the difference of real times $t-t'$, and the Fourier transformed
oscillator Green's function satisfies the matrix Dyson equation \begin{equation}
\left(D_{0}^{-1}(\omega)-\Pi(\omega)\right)D(\omega)=\hat{1}\label{DDyson}\end{equation}
 where $D_{0}^{-1}(\omega)=\left[m(\omega^{2}-\Omega_{B}^{2})\right]$,
$\Omega_{B}$ being the bare oscillator frequency, and the self-energy
(polarization operator) is a matrix of the form \begin{equation}
\Pi=\left(\begin{array}{cc}
\Pi^{R} & \Pi^{K}\\
0 & \Pi^{A}\end{array}\right).\label{bum}\end{equation}

Assuming weak interaction of the oscillator with the tunnel junction,
the self-energy can be taken to lowest order. Calculations, details
of which can be found in appendix \ref{junctionGreen}, give the following
expression for the polarization operator, \begin{equation}
\Pi(\omega)=-iG_{ww}\left(\begin{array}{cc}
\omega & 2S_{V}(\omega)\\
0 & -\omega\end{array}\right)+R_{ww}^{+}(\omega)\hat{1},\label{Greenb}\end{equation}
 where the conductance $G_{ww}$ is defined in Eq.~(\ref{cond})
and the second term, the real part of the self-energy, is given by
Eq\@.~(\ref{RB}); for $\omega$ of the order of the oscillator
frequency, $R_{ww}^{+}(\omega)$ can be replaced by a constant $R_{ww}^{+}(\omega)\approx R_{ww}^{+}(0)$.%
\footnote{We have not explicitly included terms in the self energy, that correspond
to a shift in the equilibrium position of the oscillator. The shift
can be absorbed by measuring the oscillator coordinate from the new
equilibrium position. We have to keep in mind, though, that the coordinate
independent part of the tunneling amplitude, $v_{\mathbf{rl}}$, has
to be changed accordingly.%
} The function \begin{equation}
S_{V}(\omega)=\frac{V+\omega}{2}\coth\frac{V+\omega}{2T}+\frac{V-\omega}{2}\coth\frac{V-\omega}{2T},\label{SV}\end{equation}
 where $T$ is the temperature of the junction and $V=eU$, $U$ being
the applied dc-voltage, is proportional to the well-known value of
the power spectrum of current noise of the isolated junction, see
e.g. Ref.~\onlinecite{Kog96}. Solving the Dyson equation, Eq.~(\ref{DDyson}),
the retarded and advanced oscillator Green's functions become \begin{equation}
D^{R}(\omega)=m^{-1}\frac{1}{\left(\omega+i\gamma_{e}\right)^{2}-\Omega^{2}},\quad D^{A}(\omega)=\left(D^{R}(\omega)\right)^{*},\label{Full DR DA}\end{equation}
 where $\gamma_{e}=-\Im\Pi^{R}(\omega)/2m\omega$, the damping coefficient
due to the coupling to the junction is \begin{equation}
\gamma_{e}=\frac{G_{ww}}{2m},\end{equation}
 and the renormalized oscillator frequency is \begin{equation}
\Omega^{2}=\Omega_{B}^{2}-\gamma_{e}^{2}+\frac{1}{m}R_{ww}^{+}(0).\end{equation}
 For the Keldysh component we obtain \begin{equation}
D^{K}(\omega)=\left(D^{R}(\omega)-D^{A}(\omega)\right)\frac{S_{V}(\omega)}{\omega}.\label{flucdiss1}\end{equation}

Additionally to the environment provided by the coupling to the tunnel
junction a nanomechanical oscillator is also subject to an intrinsic
environment, e.g. phonons, acting as a heat bath and leading to damping.
This additional heat bath, which we take to have the same temperature,
$T$, as the junction, can be added phenomenologically, or explicitly
by adding the interaction with a bath of harmonic oscillators (as
introduced in Refs.~\onlinecite{CleGir04, WabKhoRam05}). The total
damping coefficient for the harmonic oscillator will then be the sum
of the damping coefficients stemming from the tunnel junction, $\gamma_{e}$,
and the heat bath, $\gamma_{0}$, giving a total damping coefficient
$\gamma=\gamma_{e}+\gamma_{0}$. As a consequence of the additional
heat bath the relation between the oscillator Green's functions, Eq.~(\ref{flucdiss1}),
is modified according to \begin{equation}
D^{K}(\omega)=\left(D^{R}(\omega)-D^{A}(\omega)\right)\left(\frac{\gamma_{e}}{\gamma}\frac{S_{V}(\omega)}{\omega}+\frac{\gamma_{0}}{\gamma}\coth\frac{\omega}{2T}\right),\label{flucdiss}\end{equation}
 with the damping coefficient $\gamma_{e}$ given in Eq.~(\ref{Full DR DA}).
The relation for the Keldysh Green's function, Eq. (\ref{flucdiss}),
is characteristic of the oscillator interacting with the environment
which is in a non-equilibrium but steady state, and only in the absence
of a bias voltage it reduces to the fluctuation-dissipation relation
for an oscillator in thermal equilibrium at temperature $T$. However,
since the coupling of the oscillator to the junction is weak, $\gamma\ll\max(V,\Omega,T)$,
the oscillator spectral function is peaked at its frequency $\Omega$,
and Eq.~(\ref{flucdiss}) can be written in the standard form \begin{equation}
D^{K}(\omega)=\left(D^{R}(\omega)-D^{A}(\omega)\right)\coth\frac{\omega}{2T^{*}},\label{FlucDiss}\end{equation}
 where the temperature $T^{*}$ characterizing the stationary state
of the oscillator is given by \begin{equation}
\coth\frac{\Omega}{2T^{*}}=\left(\frac{\gamma_{e}}{\gamma}\frac{S_{V}(\Omega)}{\Omega}+\frac{\gamma_{0}}{\gamma}\coth\frac{\Omega}{2T}\right).\label{effectiveTemperature}\end{equation}
 The temperature of the oscillator, $T^{*}$, depends not only on
the environment temperature $T$, but also on the bias voltage of
the junction $V$, as well as the relative coupling strengths $\gamma_{e}/\gamma$
and $\gamma_{0}/\gamma$. For zero bias voltage the effective temperature
reduces to the environment temperature. A frequency dependent effective
temperature in the context of nanomechanical systems was also discussed
by Clerk,\cite{Cle04} as well as by Clerk and Bennett,\cite{CleBen05}
for coupling to a general environment. Clerk derived a Langevin equation
for a harmonic oscillator coupled to a fluctuating force, and obtained
a relation similar to Eq.~(\ref{effectiveTemperature}), defining
a frequency dependent effective temperature.

\section{Noise Properties of the junction \label{Noise}}

In this section we shall study how a harmonic oscillator interacting
with a tunnel junction influences the current noise in the tunnel
junction in the steady state. The noise properties of the junction
current can be obtained in perturbation theory. The noise spectrum
is specified by the current-current correlation function \begin{equation}
\left\langle \delta\hat{I}_{H}(t)\delta\hat{I}_{H}(t')\right\rangle =\left\langle \hat{I}_{H}(t)\hat{I}_{H}(t')\right\rangle -I^{2},\label{dIdI}\end{equation}
 where the current operator is given by Eq.~(\ref{current}) and
the subscript $H$ refers to an operator in the Heisenberg picture.
For calculational convenience we introduce the current-current correlator
with the time arguments lying on the Keldysh contour \begin{equation}
S(\tau,\tau')=\left\langle \mathrm{T}_{c}\left(\delta\hat{I}_{H}(\tau)\delta\hat{I}_{H}(\tau')\right)\right\rangle .\label{S(tau-tau')}\end{equation}
 It can be written in the interaction picture as \begin{equation}
S(\tau,\tau')=\left\langle \mathrm{T}_{c}\left(e^{-i\int_{c}d\tau\hat{H}_{T}(\tau)}\hat{I}(\tau)\hat{I}(\tau')\right)\right\rangle -I^{2},\label{S}\end{equation}
 where $c$ denotes the Keldysh contour. To obtain the noise spectrum
from the current-current correlator Eq.~(\ref{S}) we introduce Keldysh
indices $i,\, j=1,2$, that label the contour, e.g., $i=1$ for the
forward contour or $i=2$ for the backward contour, and revert to
using the real times $t$ and $t'$, so that \begin{equation}
S(\tau,\tau')\rightarrow S^{ij}(t-t').\label{Sij}\end{equation}
 Finally, taking the Fourier transform of Eq.~(\ref{Sij}) we obtain
\begin{equation}
S^{ij}(\omega)=\int_{-\infty}^{\infty}dt\, e^{i\omega t}S^{ij}(t).\label{Sij(omega)}\end{equation}
 For calculational purposes it is sufficient to consider the real-time
unsymmetrized current-current correlator \begin{equation}
S^{<}(t-t')=S^{12}(t-t')=\left\langle I(t)I(t')\right\rangle ,\label{cvcvc}\end{equation}
 since all other correlators can be derived from it, e.g. $S^{>}(t-t')=\left\langle I(t')I(t)\right\rangle =S^{<}(t'-t)$.

\subsection{I-V characteristic \label{Currentsection}}

First we calculate the average current $I$ to second order in the
tunneling amplitude \begin{equation}
I(t)=-i\int_{c}d\tau\left\langle \mathrm{\mathrm{T}}_{c}\left(\hat{H}_{T}(\tau)\hat{I}(t)\right)\right\rangle .\end{equation}
 Inserting the tunneling Hamiltonian Eq.~(\ref{tunneling}) and the
current operator Eq.~(\ref{current}) we get two contributions to
the current \begin{equation}
I=I_{vv}+I_{ww},\label{I2}\end{equation}
 where one part is given by the standard result for the tunnel junction
\begin{equation}
I_{vv}=G_{vv}V,\label{I-vv}\end{equation}
 with the conductance given by Eq.~(\ref{cond}).

The contribution induced by the coupling to the oscillator, $I_{ww}$,
can be written as \begin{equation}
I_{ww}=\frac{1}{2i}G_{ww}\int_{-\infty}^{\infty}d\omega J(\omega).\label{I-ww}\end{equation}
 where the conductance is given by Eq.~(\ref{cond}) and \begin{equation}
J(\omega)=VD^{K}(\omega)-\left[D^{R}(\omega)-D^{A}(\omega)\right]\Delta_{V}(\omega),\label{J(omega)}\end{equation}
 with the oscillator Green's functions found in section \ref{stationarystate},
and \begin{equation}
\Delta_{V}(\omega)=\frac{V+\omega}{2}\coth\frac{V+\omega}{2T}-\frac{V-\omega}{2}\coth\frac{V-\omega}{2T}.\label{DeltaV}\end{equation}
 Since the oscillator Green's functions are peaked at the oscillator
frequency the integral in Eq.~(\ref{I-ww}) can be evaluated and
using the expression for the Keldysh Green's functions, Eq.~(\ref{FlucDiss}),
we get \begin{equation}
I_{ww}=\frac{1}{2}\tilde{G}_{ww}\left[\Vplus\Nomega^{*}+\Vminus\left(\Nomega^{*}+1\right)\right],\label{Iwwres}\end{equation}
 where \begin{equation}
\tilde{G}_{ww}=\frac{\hbar}{m\Omega}G_{ww}\label{Gwtilde}\end{equation}
 and we introduced the short notation \begin{equation}
\Vplusminus=V\pm\Delta_{V}(\Omega)\label{Vplusminus}\end{equation}
 and the occupation number of the oscillator is given by the Bose
function \begin{equation}
\Nomega^{*}=\frac{1}{e^{\Omega/T^{*}}-1},\end{equation}
 where the temperature of the oscillator, $T^{*}$, is specified in
Eq.~(\ref{effectiveTemperature}). From Eq.~(\ref{Iwwres}) it can
be seen that two oscillator-assisted tunnelling processes contribute
to the current: tunneling electrons gaining energy from the oscillator
and loosing energy to the oscillator, respectively.

\subsection{Current-current correlator \label{currcurr}}

Next we turn to calculate the current-current correlator. For an isolated
junction, the noise is given by the second order expression in the
tunneling amplitude since the fourth order correction only introduces
a small featureless correction. However, when the junction is coupled
to a quantum system exhibiting resonant behavior, as in the case of
the oscillator, the interaction of the system with the tunnel junction
can markedly increase the fourth order noise correction at the resonance
and combinational frequencies. We will be interested in the resonant
contributions and will therefore consider the current-current correlator
to fourth order.

Expanding the expression for the current-current correlator, Eq.~(\ref{S}),
to fourth order in the tunneling amplitude we obtain \begin{equation}
S(\tau,\tau')=S_{(2)}(\tau,\tau')+S_{(4)}(\tau,\tau'),\end{equation}
 with the correlator to second order in the tunneling amplitude \begin{equation}
S_{(2)}(\tau,\tau')=\left\langle \mathrm{\mathrm{T}}_{c}\left(\hat{I}(\tau)\hat{I}(\tau')\right)\right\rangle \label{S2}\end{equation}
 and the correlator to fourth order in the tunneling amplitude \begin{multline}
S_{(4)}(\tau,\tau')=\label{S4}\\
-\frac{1}{2}\int_{c}d\tau_{1}\int_{c}d\tau_{2}\left\langle \mathrm{\mathrm{T}}_{c}\left(\hat{H}_{T}(\tau_{1})\hat{H}_{T}(\tau_{2})\hat{I}(\tau)\hat{I}(\tau')\right)\right\rangle _{L},\end{multline}
 where the subscript $L$ denotes that only linked diagrams contribute
to the expression, since the disconnected diagrams are cancelled by
the current squared term. We first investigate the second order in
tunneling contribution to the current-current correlator, resulting
in the noise floor.

\subsubsection{Noise floor}

In this section we are going to discuss the noise floor. The main
contribution to the floor comes from the second order contribution
to the noise spectrum.

To second order in the tunneling amplitudes the current-current correlator
\begin{equation}
S_{(2)}(\tau,\tau')=S_{vv}(\tau,\tau')+S_{ww}(\tau,\tau'),\end{equation}
 consists of the correlator for the isolated junction \begin{equation}
S_{vv}(\tau,\tau')=\left\langle \mathrm{T}_{c}\left(j_{v}(\tau)j_{v}(\tau')\right)\right\rangle =-i\Pi_{vv}^{+}(\tau,\tau'),\end{equation}
 and a contribution induced by the coupling to the oscillator \begin{equation}
S_{ww}(\tau,\tau')=\left\langle \mathrm{\mathrm{T}}_{c}\left(j_{w}(\tau)j_{w}(\tau')\right)\right\rangle \left\langle \mathrm{\mathrm{T}}_{c}\left(\hat{x}(\tau)\hat{x}(\tau')\right)\right\rangle ,\end{equation}
 and therefore \begin{equation}
S_{ww}(\tau,\tau')=-\Pi_{ww}^{+}(\tau,\tau')D(\tau,\tau'),\label{Sww1}\end{equation}
 where $\Pi_{vv}^{+}$ and $\Pi_{ww}^{+}$ are given by Eq.~(\ref{Pis}).
The correlator for the isolated junction can be presented as \begin{equation}
S_{vv}^{<}(\omega)=G_{vv}\left[S_{V}(\omega)-\omega\right].\label{Svvless}\end{equation}
 This result has been previously obtained by Aguado and Kouwenhoven
for the quantum point contact.\cite{AguKou00}

The symmetrized noise spectrum of an isolated junction \begin{equation}
S_{vv}^{K}(\omega)=S_{vv}^{<}(\omega)+S_{vv}^{>}(\omega),\end{equation}
 becomes the well known result \cite{DahDenLan69} \begin{equation}
S_{vv}^{K}(\omega)=2G_{vv}S_{V}(\omega),\end{equation}
 where we used the property, $S^{<}(\omega)=S^{>}(-\omega)$. In the
limit of zero voltage the noise floor reduces to Johnson-Nyquist noise.
The Markovian master equation approach employed in Ref\@.~\onlinecite{CleGir04}
and Ref.~\onlinecite{WabKhoRam05} is not able to reproduce the
correct frequency dependence, but only captures the zero frequency
noise, and gives a constant noise floor with the magnitude of $S^{K}(0)$.

At low temperatures $T\ll V$, the noise spectrum $S_{vv}^{<}$ is
controlled by the voltage $V$: at positive frequencies $\omega>|V|$
the noise is exponentially small, $S_{vv}^{<}(\omega)\approx0$, and
$S_{vv}^{<}$ increases linearly with the distance from the threshold
$\omega=|V|$: \begin{equation}
S_{vv}^{<}(\omega)=G_{vv}\left[(V-\omega)\theta(V-\omega)+(-V-\omega)\theta(-V-\omega)\right].\label{SvvT0}\end{equation}
 Eq.~(\ref{SvvT0}) shows that at $T=0$, the noise power $S_{vv}^{<}(\omega)$
is proportional to the phase space volume available for tunnelling
events, i. e., the total number of electron-hole states, with an electron
on one side of the junction and a hole on the other side, with the
excitation energy $\omega$. Obviously, this result holds only in
the frequency range where the electron density of states can be considered
as a constant.

The contribution to the noise induced by the oscillator, Eq.~(\ref{Sww1}),
becomes according Eq.~(\ref{Greenb}) \begin{equation}
S_{ww}^{<}(\omega)=2G_{ww}\int_{-\infty}^{\infty}d\omega_{1}\left[S_{V}(\omega_{1})-\omega_{1}\right]D^{<}(\omega-\omega_{1})\label{Swwless}\end{equation}
 where $D^{<}(\omega)=[D^{K}(\omega)-D^{R}(\omega)+D^{A}(\omega)]/2.$
Recalling that the oscillator Green's functions are peaked at the
oscillator frequency, the resulting contribution to the noise can
be presented as \begin{eqnarray}
S_{ww}^{<}(\omega) & = & \frac{1}{2}\tilde{G}_{ww}\left\{ \Nomega^{*}\left[S_{V}(\omega_{-})-\omega_{-}\right]\right.\nonumber \\
 &  & \left.+\left(\Nomega^{*}+1\right)\left[S_{V}(\omega_{+})-\omega_{+}\right]\right\} \label{kwe}\end{eqnarray}
 where $\omega_{\pm}=\omega\pm\Omega$ and the conductance $\tilde{G}_{ww}$
is given by Eq.~(\ref{Gwtilde}).

The second order contribution of the oscillator to the noise is similar
to that of an isolated junction: with a different coupling, $\tilde{G}_{ww}$
in the place of $G_{vv}$, the noise is given by the same expressions
apart from the frequency shift $\pm\Omega$. The two terms in Eq.~(\ref{kwe})
give the noise contribution due to processes of electron tunneling
accompanied by absorption or emission of oscillator quanta, respectively.
When the oscillator approaches the ground state, $\Nomega^{*}\rightarrow0$,
and at low junction temperatures, $T\ll\Omega,V$, the noise $S_{ww}^{<}(\omega)$
is seen to vanish for frequencies larger than $V-\Omega$.

As expected, there will be no contribution to the noise Eq.~(\ref{kwe})
for positive frequencies at zero temperature and voltage \textbf{$|V|<\Omega$},
since the voltage is insufficient to excite the oscillator.

To summarize, the second order contribution to the noise spectrum
consists of the well known noise spectrum of an isolated junction,
and a part that depends on the state of the oscillator. The second
order contribution to the noise spectrum is frequency dependent and
changes on the scale of $\max\left(V,\Omega,T\right)$.

\subsubsection{Resonant contribution to the current-current correlator }

\begin{figure}
\begin{centering}\includegraphics[width=1\columnwidth]{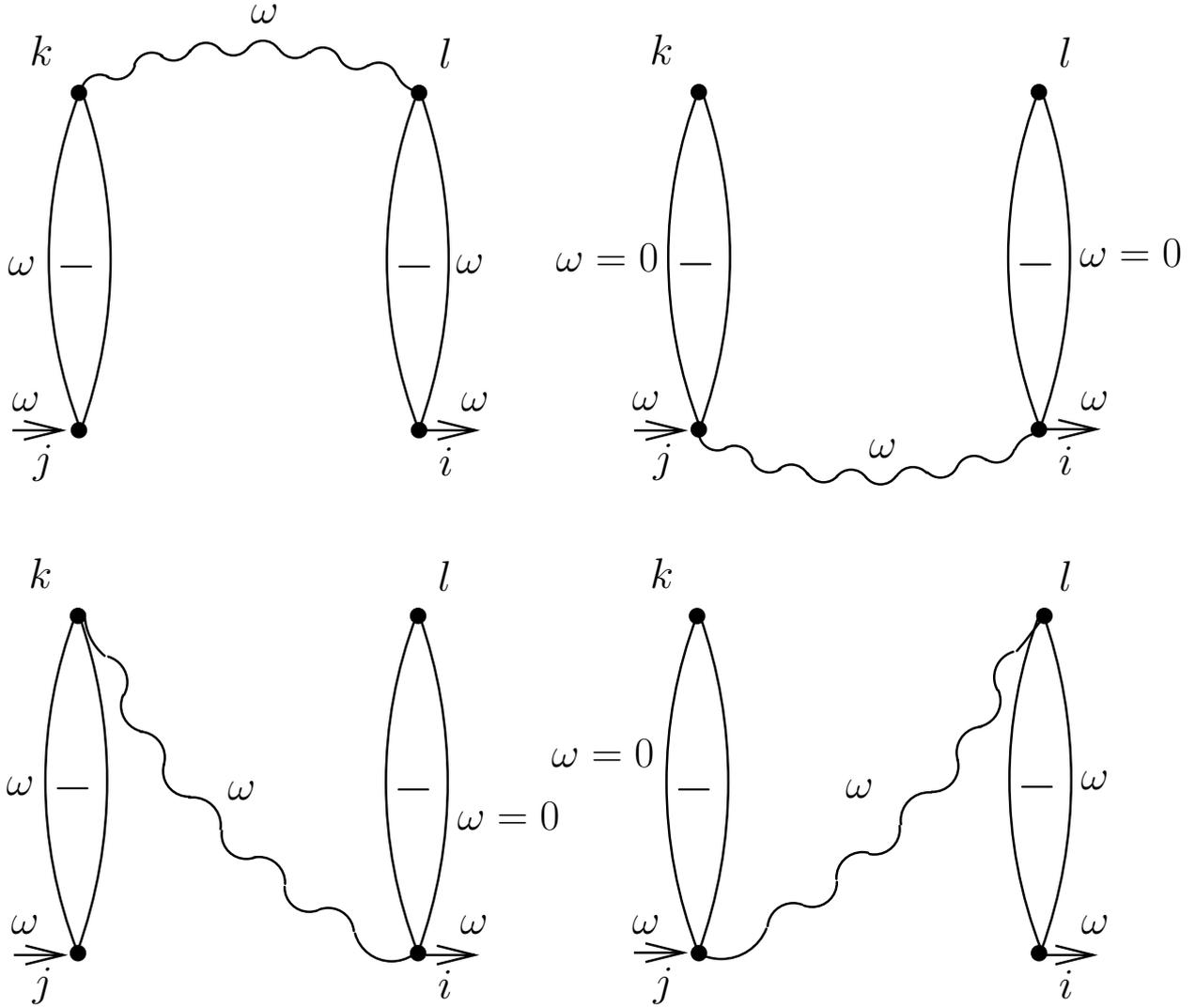}\par\end{centering}

\caption{ \label{vvwwpeak}The four diagrams proportional to $v^{2}w^{2}$
that contribute to the peak at the oscillator frequency. The wiggly
lines represent the oscillator propagator and the bubbles are defined
according to Fig.~\ref{legend2}. The short arrows labelled by $\omega$
indicate the external frequency, $i,\, j,\, k$ and $l$ are Keldysh
indices. }
\end{figure}

\begin{figure}
\begin{centering}\includegraphics[width=1\columnwidth]{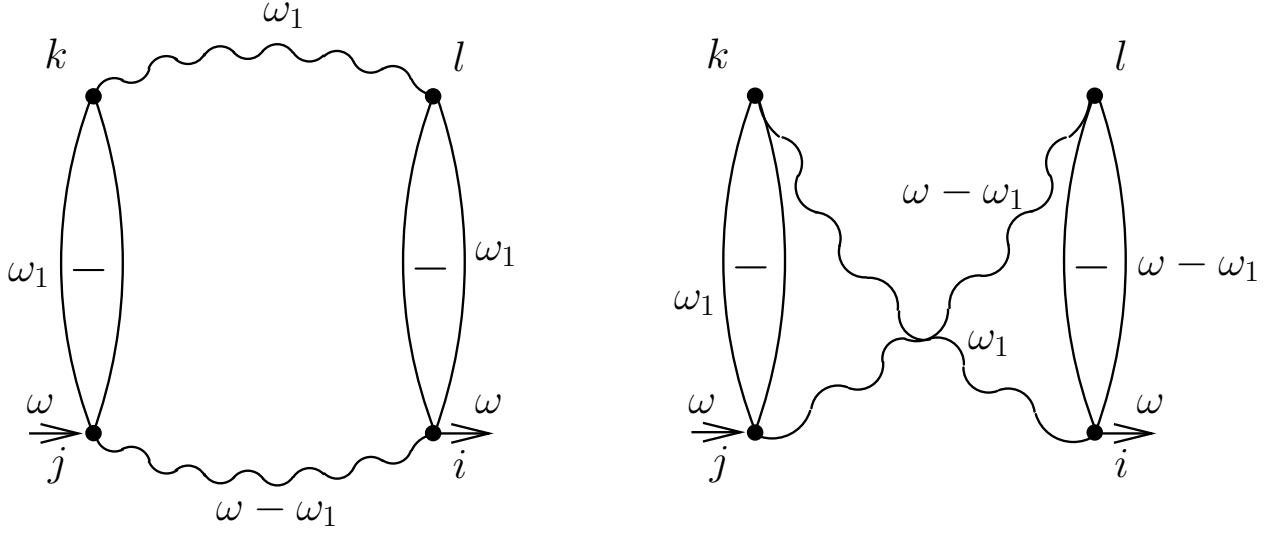}\par\end{centering}

\caption{\label{wwwwpeak}The two diagrams proportional $w^{4}$ that contribute
to the peaks at $\omega=0$ and $\omega=2\Omega$. The wiggly lines
represent the oscillator propagator and the bubbles are defined according
to Fig.~\ref{legend2}. The short arrows labelled by $\omega$ represent
the external frequency, the indices $i,\, j,\, k$ and $l$ are Keldysh
indices. Integration over the internal frequency, $\omega_{1}$ is
implicit.}
\end{figure}

\begin{figure}
\begin{centering}\includegraphics[width=0.9\columnwidth]{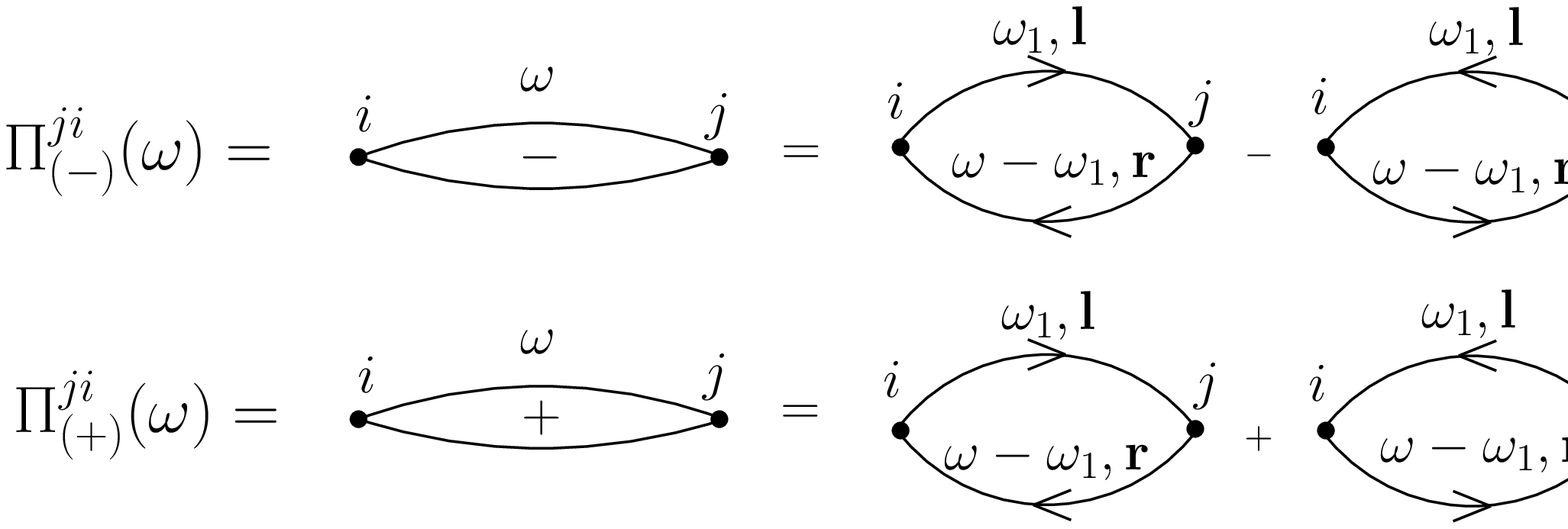}\par\end{centering}

\caption{\label{legend2}The diagrammatic representation of the junction Green's
functions, Eq.~(\ref{Pis}). Each bubble represents a linear combination
of two electron bubble diagrams, with the Keldysh indices $i$ and
$j$. Integration over the internal frequency, $\omega_{1}$, is implicit.}
\end{figure}

In this section we consider the resonant contribution to the noise
spectrum, that is, sharp peaks in the noise power with a width given
by the oscillator damping $\gamma$. Technically, the resonant contribution
originates from the fourth order terms Eq.~(\ref{S4}). Among various
Feynman diagrams generated by the expression in Eq.~(\ref{S4}),
we are therefore interested only in those that produce resonant features.
The diagrams of interest are shown in Figs.~\ref{vvwwpeak}, \ref{wwwwpeak},
where the wiggly lines represent oscillator Green's functions, while
the bubbles labelled with (--) are antisymmetric combinations of bubbles
of electron Green's functions as specified in Fig.~\ref{legend2}.

The diagrams with a single oscillator line in Fig.~\ref{vvwwpeak}
generate sharp features in the noise spectrum at $\omega=\pm\Omega$,
while the two-line diagrams in Fig.~\ref{wwwwpeak} are responsible
for noise peaks in the vicinity of $\omega=0$ and $\omega=\pm2\Omega$.

\begin{figure}
\begin{centering}\includegraphics[width=0.5\columnwidth]{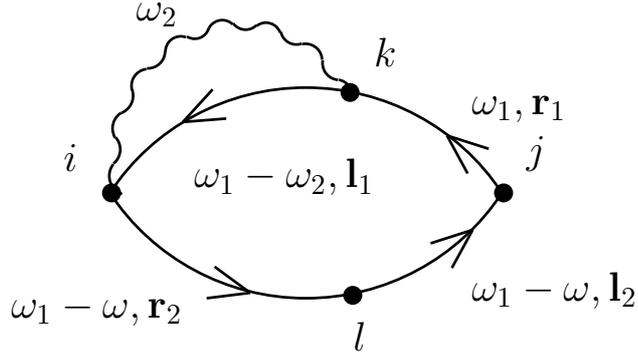}\par\end{centering}

\caption{ \label{nonbubblediagram}A diagram representing a non-bubble contribution
to the noise spectrum. The solid lines represent left and right side
electron Green's function, the wiggly line represents the oscillator
Green's functions. Integration over $\omega_{1}$ and $\omega_{2}$
is implied.}
\end{figure}

An example of a 4-th order diagram whose contribution is a featureless
function of frequency $\omega$ is shown in Fig.~\ref{nonbubblediagram}.
In this diagram, the frequency of the oscillator line is integrated
over and therefore the resonant contribution is absent. Being small
compared with the second order contribution to noise, this diagrams
and other diagrams of this type can be discarded.

We will denote the resonant contribution of the fourth order diagrams
to the current correlator by $S_{res}^{<}$. It is given by the sum
\begin{equation}
S_{res}^{<}(\omega)=S_{v^{2}w^{2}}^{<}(\omega)+S_{w^{4}}^{<}(\omega),\label{mwe}\end{equation}
 where $S_{v^{2}w^{2}}^{<}(\omega)$ and $S_{w^{4}}^{<}(\omega)$
are the contributions of the diagrams in Fig.~\ref{vvwwpeak} and
Fig.~\ref{wwwwpeak}, respectively.

Evaluating the expression represented in Fig.~\ref{vvwwpeak} by
performing the summation over the contour indices $k$ and $l$, and
inserting the bubbles and Green's functions we get (see Appendix \ref{fourthorderappendix}
for details)

\begin{equation}
S_{v^{2}w^{2}}^{<}(\omega)=-2iG_{vw}^{2}VJ(\omega),\label{SvvwwRes}\end{equation}
 where the function $J(\omega)$ is specified in Eq.~(\ref{J(omega)})
and the conductance given in Eq.~(\ref{cond}). This contribution
to the noise spectrum is peaked at $\omega=\pm\Omega$ due to the
resonant behavior of the harmonic oscillator spectral function. The
dependence of the peak height on the system parameters will be discussed
in section \ref{resonantDisc}.

The contribution containing two oscillator Green's functions in Fig.
\ref{wwwwpeak} is calculated in Appendix \ref{fourthorderappendix}
to give \begin{eqnarray}
S_{w^{4}}^{<}(\omega) & = & \frac{G_{ww}^{2}}{2}\int_{-\infty}^{\infty}d\omega_{1}J(\omega_{1})J(\omega-\omega_{1})\nonumber \\
 & + & \frac{G_{ww}^{2}}{2}\int_{-\infty}^{\infty}d\omega_{1}A(\omega_{1})A(\omega-\omega_{1})\nonumber \\
 & \times & \left[V-\Delta_{V}(\omega_{1})\right]\left[V-\Delta_{V}(\omega-\omega_{1})\right]\label{SwwwwRes}\end{eqnarray}
 where we introduced the oscillator spectral function $A(\omega)=i\left[D^{R}(\omega)-D^{A}(\omega)\right].$
The dependence of the contribution to the noise spectrum, Eq.~(\ref{SwwwwRes}),
on system parameters will be discussed in section \ref{resonantDisc}.

\subsection{Resonant contributions to the noise spectrum \label{resonantDisc}}

The resonant contribution to the noise consists of peaks at zero frequency,
$\omega=0$, the oscillator frequency, $\omega=\pm\Omega$, and twice
the oscillator frequency, $\omega=\pm2\Omega$. Their heights depend
inversely on the damping coefficient of the oscillator, $\gamma$,
and their width is proportional to the damping coefficient. In this
section we will discuss the properties of the peaked structure and
its dependence on environment parameters like bias voltage and temperature.
The peaked contribution to the noise can be written as \begin{equation}
S_{res}^{<}(\omega)=S_{-2}^{<}(\omega)+S_{-1}^{<}(\omega)+S_{0}^{<}(\omega)+S_{1}^{<}(\omega)+S_{2}^{<}(\omega),\end{equation}
 where $S_{0}^{<}(\omega)$ describes the peaked contribution to the
noise spectrum at zero frequency, $S_{\pm1}^{<}(\omega)$ the contributions
at the oscillator frequency and $S_{\pm2}^{<}(\omega)$ the contributions
at twice the oscillator frequency.

The peaks at the frequencies $\omega=\pm\Omega$ originate from the
bubble contribution, $S_{v^{2}w^{2}}^{<}(\omega)$, given in Eq.~(\ref{SvvwwRes}).
Inserting the explicit form of the oscillator Green's functions, Eqs.~(\ref{Full DR DA},
\ref{FlucDiss}), in Eq.~(\ref{SvvwwRes}), and using that the oscillator
Green's functions are peaked at the oscillator frequency\textbf{,}
we obtain for the resonant contribution to the noise spectrum at the
oscillator frequency \begin{equation}
S_{\pm1}^{<}(\omega)=\frac{\gamma^{2}}{\left(\omega\mp\Omega\right)^{2}+\gamma^{2}}P_{\Omega}.\end{equation}
 The peak height at the resonance frequency is \begin{equation}
P_{\Omega}=\frac{2}{\gamma}\tilde{G}_{vw}^{2}V\left[\Nomega^{*}\Vplus+\left(\Nomega^{*}+1\right)\Vminus\right],\label{PeakHeightOmega}\end{equation}
 where $\Vplusminus$ is given by Eq.~(\ref{Vplusminus}) and $\Nomega^{*}$
is the occupation number of the oscillator. The peak height scales
inversely with the damping coefficient $\gamma$. For large voltages
$V\gg\Omega$ the peak height is linear in the occupation number $\Nomega^{*}$
and quadratic in the voltage. The result, Eq.~(\ref{PeakHeightOmega}),
extends the result from the Markovian master equation calculation
in Refs.~\onlinecite{CleGir04, WabKhoRam05} to voltages smaller
than the oscillator frequency.

The peaks at frequencies $\omega=0$ and $\omega=\pm2\Omega$ come
from the fourth order contribution quadratic in the oscillator Green's
functions, Eq.~(\ref{SwwwwRes}). The remaining integration in Eq.~(\ref{SwwwwRes})
can be done and the dominating contribution for weak damping comes
from the poles of the oscillator Green's functions. Collecting the
contributions to the peak at zero frequency we get \begin{equation}
S_{0}^{<}(\omega)=\frac{4\gamma^{2}}{\omega^{2}+4\gamma^{2}}P_{0}\label{6we}\end{equation}
 where the peak height $P_{0}$ is \begin{equation}
P_{0}=\frac{1}{2\gamma}\tilde{G}_{ww}^{2}V\left[\Nomega^{*2}\Vplus+(\Nomega^{*}+1)^{2}\Vminus\right].\label{PeakHeightZero}\end{equation}
 The peak height at zero frequency also scales inversely with the
damping coefficient $\gamma$ and depends quadratically on the conductance
$\tilde{G}_{ww}$. The result for the peak at zero frequency, Eq.~(\ref{6we}),
coincides with the result obtained by using a Markovian master equation
approach as in Ref.~\onlinecite{WabKhoRam05}, which therefore
correctly captures the low frequency noise.

The contributions at double the oscillator frequency differ for positive
and negative frequency. In the vicinity of $\omega=2\Omega$, \begin{equation}
S_{2}^{<}(\omega)=\frac{4\gamma^{2}}{\left(\omega-2\Omega\right)^{2}+4\gamma^{2}}P_{2\Omega}\end{equation}
 with the peak height given by \begin{equation}
P_{2\Omega}=\tilde{G}_{ww}^{2}\frac{1}{4\gamma}V\Nomega^{*}\left[\Vplus\Nomega^{*}+\Vminus\left(\Nomega^{*}+2\right)\right]\label{PeakHeight2Om}\end{equation}
 and for $\omega\sim-2\Omega$ \begin{equation}
S_{-2}^{<}(\omega)=\frac{4\gamma^{2}}{\left(\omega+2\Omega\right)^{2}+4\gamma^{2}}P_{-2\Omega}\end{equation}
 with the peak height \begin{eqnarray}
P_{-2\Omega} & = & \tilde{G}_{ww}^{2}\frac{1}{4\gamma}V\left(\Nomega^{*}+1\right)\nonumber \\
 &  & \times\left[\Vplus\left(\Nomega^{*}-1\right)+\Vminus\left(\Nomega^{*}+1\right)\right].\end{eqnarray}
 The peak\textbf{s} at double the oscillator frequency also show resonant
behavior, i.e. the peak height increases with decreasing damping.
For large voltages and occupation number, $\Nomega^{*}\gg1$, it depends
quadratically on the oscillator occupation number and the voltage.

The peak heights depend on the bias $V$, the environment temperature
$T$ and the relative coupling strengths $\gamma_{e}/\gamma$ and
$\gamma_{0}/\gamma$. In the following we discuss the peak heights
in some limiting cases.

\subsubsection{Dominant coupling to the tunnel junction}

If the coupling between the oscillator and the tunnel junction is
much stronger than the coupling to the thermal environment, $\gamma_{e}\gg\gamma_{0}$,
the occupation number of the oscillator is, according to Eq.~(\ref{effectiveTemperature}),
given by \begin{equation}
\Nomega^{*}=\frac{1}{2}\left(\frac{S_{V}(\Omega)}{\Omega}-1\right).\end{equation}

\paragraph{For high temperatures $T\gg V,\Omega$ }

the occupation number of the oscillator becomes $\Nomega^{*}\approx T/\Omega$.
The functions $\Vplusminus=V$ in this limit. The peak heights of
the resonant contributions to the noise spectrum are then at the oscillator
frequency \begin{equation}
P_{\Omega}=\tilde{G}_{vw}^{2}\frac{4}{\gamma}V^{2}\frac{T}{\Omega},\label{S1T}\end{equation}
 and is linear in the environment temperature. The peak height at
zero frequency is

\begin{eqnarray}
P_{0} & = & \tilde{G}_{ww}^{2}\frac{1}{\gamma}V^{2}\left(\frac{T}{\Omega}\right)^{2},\label{S0T}\end{eqnarray}
 and depends quadratically on the environment temperature, whereas
the peak heights at twice the oscillator frequency become \begin{equation}
P_{\pm2\Omega}=\tilde{G}_{ww}^{2}\frac{1}{2\gamma}V^{2}\left(\frac{T}{\Omega}\right)^{2},\label{S2T}\end{equation}
 which also scales quadratically in the environment temperature. For
dominant coupling to the junction $\gamma\propto G_{ww}$ and therefore
all the peak heights becomes linear functions of the conductance.
The peak heights for high voltages $V\gg T,\Omega$ can be obtained
from the peak heights at high temperatures by replacing $T\rightarrow V/2$
in Eqs.~(\ref{S1T}, \ref{S0T}, \ref{S2T}).

\paragraph{At low temperatures, $T\ll\Omega$, }

and low voltages $V<\Omega$, the occupation number approaches zero,
$\Nomega^{*}\approx0$. The peaks at the oscillator frequency as well
as the peak at zero frequency and the peak at $\omega=2\Omega$ disappear
\begin{equation}
P_{0}=P_{\Omega}=P_{2\Omega}=0.\label{Peaksarezero}\end{equation}
 At $\omega=-2\Omega$ we obtain a dip in the noise spectrum with
depth \[
P_{-2\Omega}=-\tilde{G}_{ww}^{2}\frac{1}{2\gamma}V^{2}.\]
 Since in nanomechanical systems the coupling strength between the
junction and the oscillator can be tuned, we will also discuss the
case where the thermal environment dominates over the junction.

\subsubsection{Dominant coupling to the thermal environment }

Another limiting case to consider is the situation when the coupling
to the thermal environment dominates over the coupling to the tunnel
junction, $\gamma_{0}\gg\gamma_{e}$. In the high temperature limit,
$T\gg V,\Omega$, we obtain the same results for the peak heights
as in the case of dominant coupling to the junction, since for both
cases the junction and the thermal environment act as thermal equilibrium
environments with temperature $T$. For high voltages and temperatures
much larger than the oscillator frequency, $V\gg T\gg\Omega$, we
get \begin{equation}
\Nomega^{*}=\frac{\gamma_{e}}{\gamma}\frac{|V|}{2\Omega}+\frac{\gamma_{0}}{\gamma}\frac{T}{\Omega}\gg1.\end{equation}
 The peak height at the oscillator frequency is \begin{equation}
P_{\Omega}=\tilde{G}_{vw}^{2}\frac{4}{\gamma}V^{2}\left(\frac{\gamma_{e}}{\gamma}\frac{|V|}{2\Omega}+\frac{\gamma_{0}}{\gamma}\frac{T}{\Omega}\right).\end{equation}
 The peak height at zero frequency is \begin{equation}
P_{0}=\tilde{G}_{ww}^{2}\frac{1}{\gamma}V^{2}\left(\frac{\gamma_{e}}{\gamma}\frac{|V|}{2\Omega}+\frac{\gamma_{0}}{\gamma}\frac{T}{\Omega}\right)^{2}.\end{equation}
 The peak height at twice the oscillator frequency \begin{equation}
P_{\pm2\Omega}=\tilde{G}_{ww}^{2}\frac{1}{2\gamma}V^{2}\left(\frac{\gamma_{e}}{\gamma}\frac{|V|}{2\Omega}+\frac{\gamma_{0}}{\gamma}\frac{T}{\Omega}\right)^{2}\end{equation}
 is also quadratic in the occupation number. All peak heights depend
inversely on the damping coefficient $\gamma$, since in the parenthesis
$\gamma$ appears only in the relative coupling strengths $\gamma_{e}/\gamma$
and $\gamma_{0}/\gamma$.

For low temperatures $T\ll V,\Omega$ we distinguish two regimes.
For low voltages, $V<\Omega$, we obtain the same results as in the
case for strong coupling to the junction since $\Nomega^{*}\approx0$.
For voltages $V\gg\Omega$ \begin{equation}
\Nomega^{*}=\frac{\gamma_{e}}{\gamma}\frac{|V|}{2\Omega}.\end{equation}
 The functions $\Vplusminus\approx V\pm\Omega$ and we obtain for
the peak height at the oscillator frequency \begin{equation}
P_{\Omega}=\tilde{G}_{vw}^{2}\frac{2}{\gamma}V^{2}\left(2\left(\frac{\gamma_{e}}{\gamma}\frac{|V|}{2\Omega}\right)+1\right),\end{equation}
 for the peak height at zero frequency \begin{equation}
P_{0}=\tilde{G}_{ww}^{2}\frac{1}{2\gamma}V^{2}\left[2\left(\frac{\gamma_{e}}{\gamma}\frac{V}{2\Omega}\right)^{2}+2\left(\frac{\gamma_{e}}{\gamma}\frac{|V|}{2\Omega}\right)+1\right]\end{equation}
 and the peak heights at twice the oscillator frequency \begin{equation}
P_{2\Omega}=\tilde{G}_{ww}^{2}\frac{1}{2\gamma}V^{2}\left(\frac{\gamma_{e}}{\gamma}\frac{|V|}{2\Omega}\right)\left[\left(\frac{\gamma_{e}}{\gamma}\frac{|V|}{2\Omega}\right)+1\right]\end{equation}
 and \begin{equation}
P_{-2\Omega}=\tilde{G}_{ww}^{2}\frac{1}{2\gamma}V\left[\left(\frac{\gamma_{e}}{\gamma}\frac{|V|}{2\Omega}\right)+1\right]\left[V\left(\frac{\gamma_{e}}{\gamma}\frac{|V|}{2\Omega}\right)-\Omega\right].\end{equation}
 We have obtained the peak heights of the resonant peaks in the noise
spectrum of a tunnel junction coupled to a harmonic oscillator for
arbitrary parameters $V,\, T$ and $\gamma_{e},\gamma_{0}$. We find
that if the oscillator approaches the ground state, $\Nomega^{*}\rightarrow0$,
the peaks at positive frequencies as well as the peak at zero vanish.
In the next section we are going to discuss an application of the
properties of the peaks in the noise spectrum of the junction: noise
thermometry.

\section{Noise Thermometry \label{thermometry}}

In experiments trying to cool a nanomechanical oscillator to the ground
state a diagnostic tool is needed to check the state of the oscillator
and to confirm, eventually, that the oscillator really is in the ground
state. Coupling the oscillator to an electrical device, e.g. a tunnel
junction, gives a means to experimentally determine the state of the
oscillator, since the oscillator influences the current and noise
in the junction. In a noise thermometry setup the noise that the oscillator
induces in the tunnel junction is used to determine the temperature
or occupation number of the oscillator.

\begin{figure}
\begin{centering}\includegraphics[width=0.9\columnwidth]{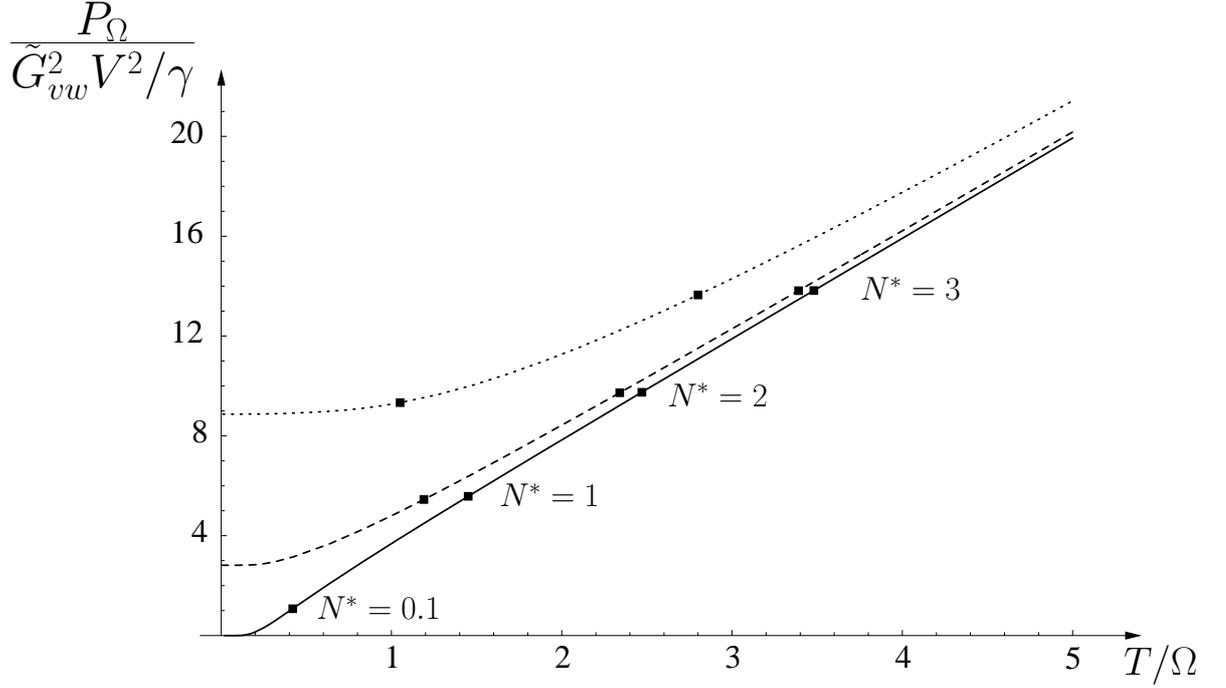}\par\end{centering}

\caption{ \label{S1ofT}The peak height of the noise at the oscillator frequency,
$P_{\Omega}$, as a function of the environment temperature $T$ for
different bias voltages and dominant coupling to the junction, $\gamma_{e}/\gamma_{0}=10$,
and $\tilde{G}_{vw}=0.1$. The full line shows the temperature dependence
of the peak height for $V=0.1\Omega$, the dashed line for $V=2\Omega$
and the dotted line for $V=5\Omega$. As the environment temperature
approaches zero, the peak height vanishes for voltages $V<\Omega$,
but stays finite for voltages $V>\Omega$. The square dots mark different
occupation numbers of the oscillator as indicated.}
\end{figure}

As demonstrated, the noise spectrum of a tunnel junction consists
of three peaks, one at zero frequency, one at the oscillator frequency
and one at twice the oscillator frequency, and the peak heights depend
on the oscillator occupation number. If the oscillator couples weakly
to the junction the highest peak is at the oscillator frequency $\Omega$.
The relation for the peak height, Eq.~(\ref{PeakHeightOmega}), can
be used to determine the oscillator occupation number in an experiment.
The usual experimental procedure (see Refs.~\onlinecite{NaiBuuLah06,
LahBuuCam04}) is to measure the peak height as a function of temperature.
Fig\@.~\ref{S1ofT} shows the expected outcome of such a measurement,
the dependence of the peak height at the oscillator frequency, obtained
from Eq.~(\ref{PeakHeightOmega}), as a function of the environment
temperature $T$ for different bias voltages $V$ and dominant coupling
to the junction $\gamma_{e}/\gamma=10$. The solid line shows the
peak height for $V=0.1\Omega$. It decreases monotonically for decreasing
temperature, and becomes exponentially small at low temperatures.
The dashed and the dotted line show the peak height for voltages larger
than the oscillator frequency ($V=2\Omega$ and $V=5\Omega$ respectively).%

The height of the peak $P_{\Omega}$ can be conveniently scaled to
a dimensionless quantity $N_{c}$, \begin{equation}
N_{c}=\gamma\frac{P_{\Omega}}{4\tilde{G}_{vw}^{2}V^{2}},\label{56hgf}\end{equation}
 where the combination $\gamma/\tilde{G}_{vw}^{2}$ is a property
of the device. For high occupation numbers (i.e. high temperatures
or high voltage, $\max(V,T)\gg\Omega$), the dimensionless peak height
$N_{c}$ gives the oscillator occupation number: $\Nomega^{*}\approx N_{c}$.
The normalization constant, $4\tilde{G}_{vw}^{2}V^{2}/\gamma$ in
Eq.~(\ref{56hgf} ), can be read off as the high temperature slope
in a plot of $P_{\Omega}$ vs. $T/\Omega$. In the region of low occupation
numbers we obtain the relation for the occupation number \begin{equation}
\Nomega^{*}=N_{c}-\frac{1}{2}\frac{\Vminus}{V}.\label{98df}\end{equation}
 where $\Vminus$, Eq.~(\ref{Vplusminus}), is a known function of
the bias and temperature.

\begin{figure}
\noindent \begin{centering}\includegraphics[width=0.7\columnwidth,keepaspectratio]{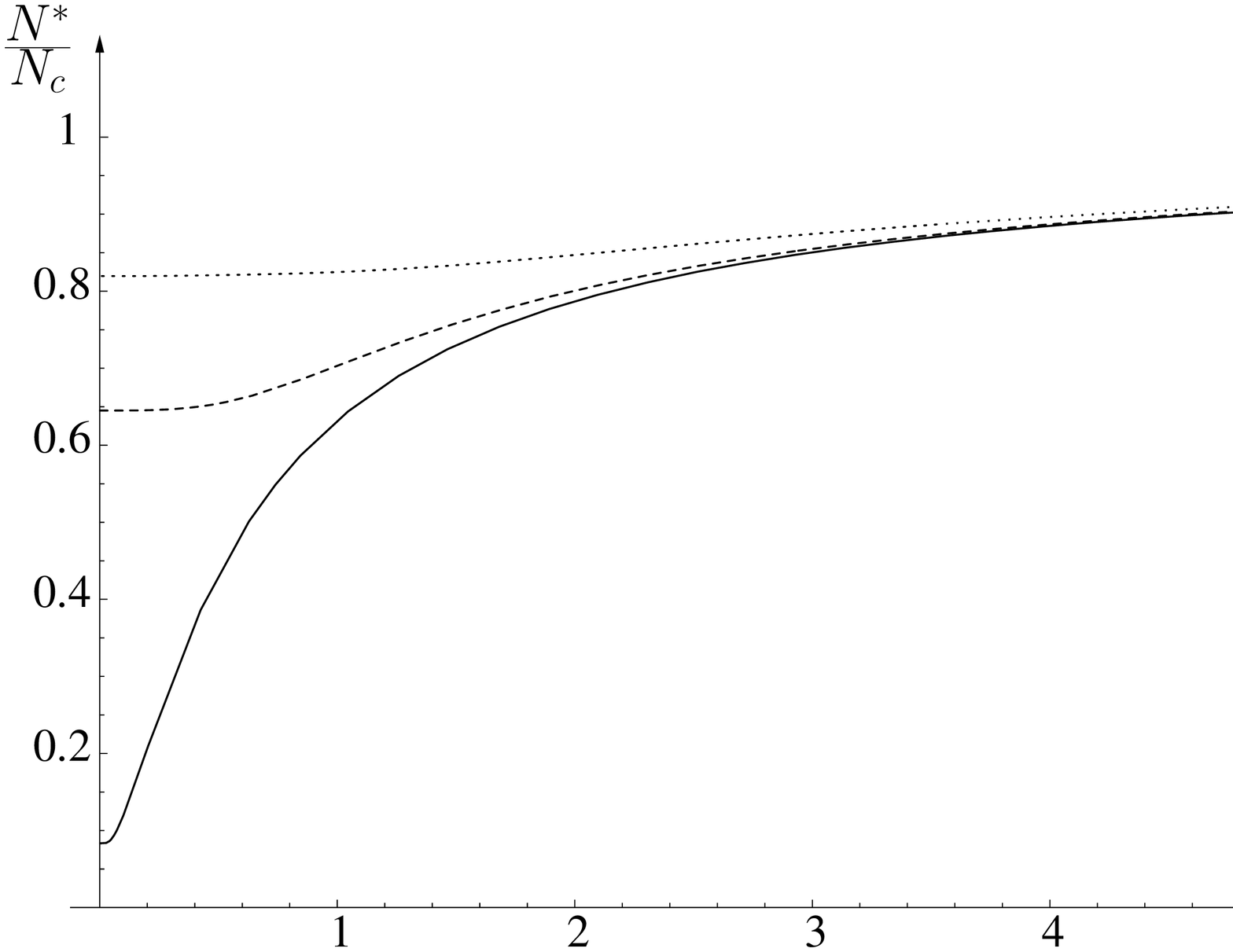}\par\end{centering}

\caption{ \label{SoverN}The relative occupation number of the oscillator
$\Nomega^{*}/N_{c}$ as a function of the environment temperature
$T$ for different bias voltages $V$ and dominant coupling to the
junction, $\gamma_{e}/\gamma_{0}=10$, with $\tilde{G}_{vw}=0.1$.
The full line shows the relative occupation number for low voltage,
$V=0.1\Omega$, the dashed line and the dotted line show the relative
occupation number for high voltages, $V=2\Omega$ and $V=5\Omega$,
respectively. For high temperatures $T\gg\Omega$, the relative occupation
number approaches unity. For low temperatures the relative occupation
number can depart considerably from unity to approach $1/(1+\gamma\Omega/\gamma_{e}V)$.}
\end{figure}

We note that for low temperatures and voltages the oscillator occupation
number is not simply proportional to the peak height at the oscillator
frequency. The relative occupation number $\Nomega^{*}/N_{c}$ is
shown in Fig.~\ref{SoverN} for dominant coupling to the junction
and different bias voltages. At high temperatures the relative occupation
number approaches unity, for low temperatures and low voltages it
departs from unity.%

Another possibility to get information on the oscillator occupation
number is from the peak heights at $\omega=0$ and $\omega=2\Omega$
using Eqs.~(\ref{PeakHeightZero}, \ref{PeakHeight2Om}). The combination
of junction parameters $V^{2}\tilde{G}_{ww}^{2}/\gamma$ in Eqs.~(\ref{PeakHeightZero},
\ref{PeakHeight2Om}) can be obtained, e.g, from the high temperature
slope in a plot of $P_{0}$ vs. $T^{2}/\Omega^{2}$.

\begin{figure}
\begin{centering}\includegraphics[width=1\columnwidth]{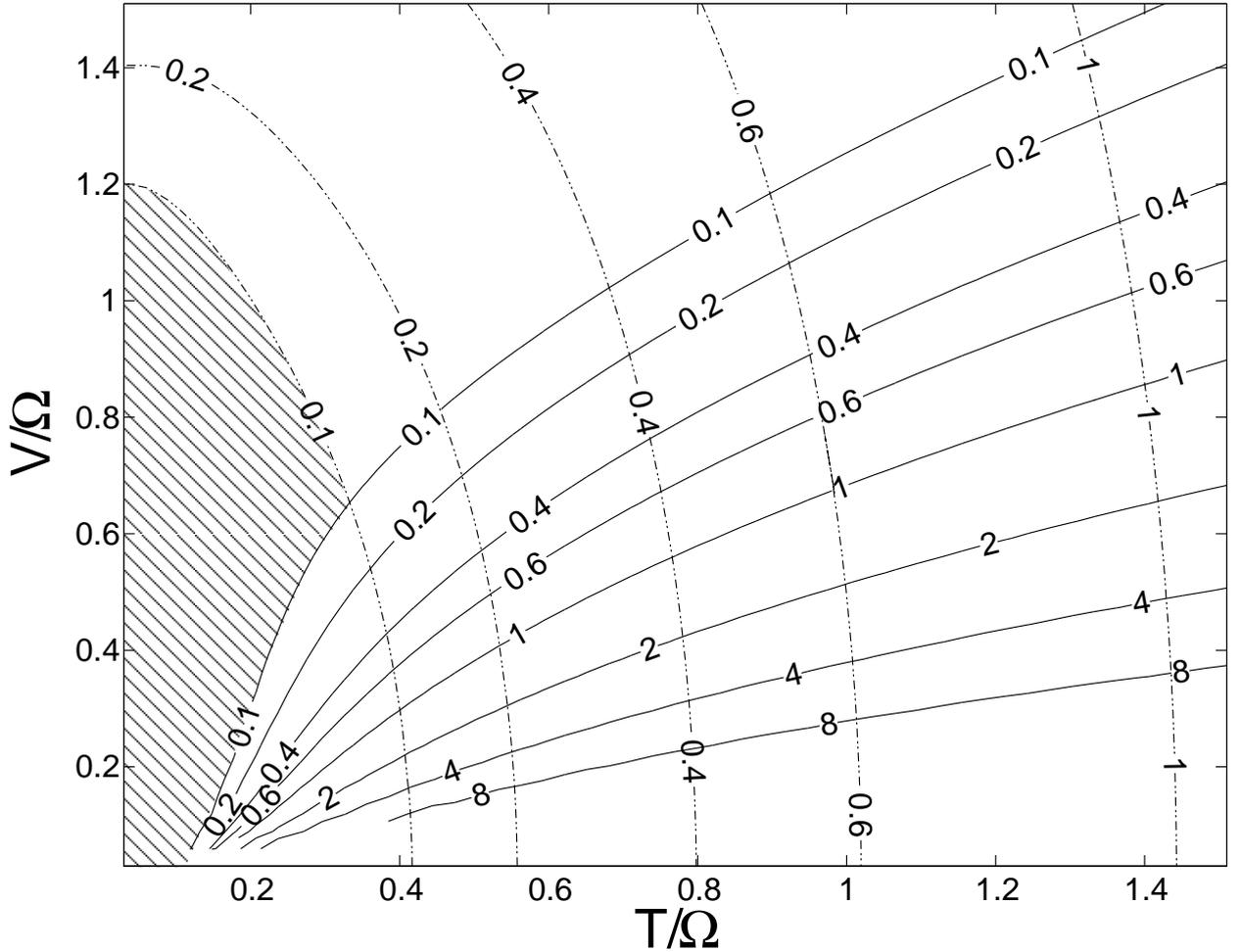}\par\end{centering}

\caption{\label{sn}The solid lines show are a contour plot of the minimum
occupation number necessary to observe the peak at the oscillator
frequency with a S/N ratio $r>0.2$, as a function of voltage and
temperature with relative conductances $G_{vv}/\tilde{G}_{vw}=10$
and $G_{ww}/\tilde{G}_{vw}=0.1$ . The dashed lines are a contour
plot of the occupation number of the oscillator for dominant coupling
to the junction, $\gamma_{e}/\gamma_{0}=100$. The shaded region to
the left marks the parameter regime that allows a readout of the oscillator
occupation number without heating the oscillator.}
\end{figure}

Since the peak at the oscillator frequency is not isolated, but sits
on a noise floor, the question of observability of the peak arises.
As a measure of observability we define a signal to noise ratio, the
peak height relative to the floor \begin{equation}
r=\frac{P_{\Omega}}{S_{vv}(\Omega)+S_{ww}(\Omega)},\label{snratio}\end{equation}
 and demand $r>0.1$ for the peak to be observable. The peak height,
$P_{\Omega}$, as well as the magnitude of the noise floor, $S_{vv}(\Omega)+S_{ww}(\Omega)$,
depend on $N^{*}$, the junction bias, and the junction temperature
as well as the conductances $G_{vv},\,\tilde{G}_{vw}$ and $\tilde{G}_{ww}$.
To investigate the observability of the peak, we use Eq.~(\ref{snratio})
to find the occupation number $N^{*}$ for a given bias voltage, junction
temperature and a set of conductances, so that $r=0.2$. The solid
lines in Fig.~\ref{sn} show the occupation number necessary to fulfill
$r=0.1$ as a function of voltage and junction temperature. To detect
an oscillator close to its ground state high bias voltages and / or
low junction temperatures are necessary, as seen in Fig.~\ref{sn}.
On the other hand, high voltages heat the oscillator, so that a compromise
between heating and readout has to be found. To illustrate the heating
effect, the dash-dotted lines show the occupation number for dominant
coupling to the junction, $\gamma_{e}/\gamma_{0}=100$. If one would
like to read out the oscillator occupation number when $N^{*}<0.1$,
without introducing heating, voltage and temperature are confined
to the shaded region on the left of Fig.~\ref{sn}. This example
shows that a good signal to noise ratio for the readout and a minimal
heating of the oscillator are complimentary requirements. %

\section{Conclusion \label{conclusions}}

We have considered a nanomechanical resonator interacting with a dc-biased
tunnel junction. We model the resonator as a harmonic oscillator and
the interaction is introduced via the modulation of the tunneling
amplitude by coupling to the harmonic oscillator position. Employing
a Green's function technique we calculated the properties of the stationary
state of the oscillator, obtaining that the coupling to the junction
introduces damping and heats the oscillator. The expression Eq.~(\ref{effectiveTemperature}),
for the temperature of a harmonic oscillator coupled to a tunnel junction
was also obtained in Ref.~\onlinecite{WabKhoRam05} using a Markovian
master equation approach. The reason for the coinciding results derived
by two different techniques is due to the same weak coupling approximation
made in both derivations. Both, the perturbative Green's function
calculation presented here and the Markovian master equation can only
be applied for weak coupling between the environment and the oscillator.
In the master equation approach the weak coupling leads to a separation
of time scales of the environment evolution and the evolution of the
density matrix, resulting in a Markovian master equation. Here the
same argument is used in the frequency domain: The width of the oscillator
spectral function is small on the scale of the frequency dependence
of the environment correlation functions.

The current-voltage characteristic of the junction has been obtained.
The stationary state current is seen to consist of two contributions.
One contribution is the current through an isolated junction, whereas
the other contribution, dependent on the state of the oscillator,
stems from oscillator assisted tunneling. The expression obtained
for the dc-current is in accordance with the result previously derived
using a master equation technique.\cite{WabKhoRam05,bum} The additional
current arising from the influence of the oscillator on the junction
transmission, Eq.~(\ref{Iwwres}), vanishes if the oscillator is
in the ground state and the voltage across the junction is smaller
than the oscillator frequency. We observe therefore that the zero
point fluctuations of the oscillator position do not affect the electric
current.

The main part of the paper presents the calculation of the electric
noise due to oscillator-assisted tunnelling. The unsymmetrized current-current
correlator has been evaluated for arbitrary frequencies, bias voltages,
and environment temperatures. The noise spectrum consists of a smooth
noise floor and a peaked resonant structure.

As a function of frequency $\omega$, the noise floor varies on the
scale $\max(T,V,\Omega)$. The contribution of the oscillator is shown
to vanish at positive frequencies when the oscillator is in the ground
state and the bias voltage is smaller than the oscillator frequency.
The expressions Eqs.~(\ref{Svvless}, \ref{kwe}) for the second
order noise generalize the result obtained by a Markovian master equation
derived in Ref.~\onlinecite{WabKhoRam05}. In the Markovian master
equation calculation a constant noise floor was obtained, whereas
the Green's function calculation presented here captures the frequency
dependent noise floor. As expected, in the low frequency limit, we
recover the result obtained by the Markovian master equation approach.

The resonant structure in the noise spectrum, $S^{<}(\omega)$, consists
of peaks at zero frequency, $\omega=0$, the oscillator frequency,
$\omega=\pm\Omega$, and twice the oscillator frequency, $\omega=\pm2\Omega$.
Being present only for a finite voltage across the junction, the resonances
are of non-equilibrium origin, and their intensities at positive and
negative frequencies are not related by the detailed balance relation.
\cite{GavLevImr00} The peaks at the oscillator frequency have the
same height for positive and negative frequencies, whereas the peaks
at double the oscillator frequency are asymmetric. The peaks at $\omega=\pm\Omega$
stem from processes involving a single vibrational quantum, and their
height is therefore linear in the oscillator occupation number $N^{*}$,
whereas the processes leading to the peaks at $\omega=0$ and $\omega=\pm2\Omega$
involve two quanta and the peak heights are quadratic in $N^{*}$.
If the oscillator is in the ground state, the peaks at positive frequencies
vanish in the limit where the bias voltage is smaller than the oscillator
frequency but a dip in the noise spectrum remains for negative frequencies.

For bias voltages smaller than the oscillator frequency and the oscillator
in the ground state, we find no oscillator dependent noise at positive
frequencies: Both the oscillator dependent contribution to the noise
floor as well as the peaks at positive frequencies disappear from
the unsymmetrized noise spectrum, $S^{<}(\omega)$. The oscillator
contribution to the current-current correlator at negative frequencies
remains finite even for voltages smaller than the oscillator frequency.

To understand this result we consider a setup for measuring current
fluctuations. Lesovik and Loosen\cite{LesLoo97} as well as Gavish,
Levinson and Imry,\cite{GavLevImr00} introduced a damped harmonic
oscillator with resonance frequency $\Omega_{0}$ as a meter of current
fluctuations. For linear coupling of the oscillator to the current
in the junction, they obtained the deviation of the meter position
fluctuations, $\langle X^{2}\rangle$, from the equilibrium value
in terms of the current-current correlators of the junction, $S^{<}$
and $S^{>}$ (see Eq.~(\ref{cvcvc})), and the meter occupation number,
$N_{\Omega_{0}}$, \begin{equation}
\delta\left\langle \hat{X}^{2}\right\rangle =A\left[\left(N_{\Omega_{0}}+1\right)S^{<}(\Omega_{0})+N_{\Omega_{0}}S^{>}(\Omega_{0})\right],\label{meterFluctuations}\end{equation}
 where $A$ is the coupling strength and $N_{\Omega_{0}}$ is the
Bose function with temperature $T$. Note that the argument of the
noise spectrum $S^{<,>}$, the oscillator frequency $\Omega_{0}$
is positive. Let us consider the meter fluctuations in different temperature
ranges. For large detector temperatures $T\gg\Omega_{0}$, and consequently
$N_{\Omega_{0}}\gg1$, the meter fluctuations are proportional to
the symmetrized current-current correlator of the junction, $S^{>}+S^{<}$.
A \emph{passive} detector -- a detector at low temperature $T\ll\Omega_{0}$,
when $N_{\Omega_{0}}\ll1$ -- measures the unsymmetrized current-current
correlator $S^{<}$ at \emph{positive} frequency. In accordance with
these arguments,\cite{LesLoo97,GavLevImr00} only the part of $S^{<}$
at positive frequencies describes physical noise, understood as random
flow of energy from the system to environment.

We can now understand our results for the current-current correlator
$S^{<}(\omega)$ with the help of Eq.~(\ref{meterFluctuations}).
A passive detector detects only the positive frequency part of the
current-current correlator. Our calculations show vanishing $S^{<}(\omega)$
at positive frequencies $\omega$ in the case when the oscillator
is in the ground state, and thus the ground state does not contribute
to the physical noise in the tunnel junction in compliance with general
expectations.

In experiments with nanomechanical resonators utilizing electrical
devices as detectors, noise properties can function as a diagnostic
tool in determining the state of the resonator. We have shown how
peaks in the noise power spectrum can act as a measure for the oscillator
occupation number and discussed the criteria for observing these peaks.

In this paper we presented a calculation of the noise induced by an
oscillator in a tunnel junction. The obtained results are valid at
arbitrary parameters and thus also in the important region where the
oscillator approaches the ground state.

\appendix

\section{\label{junctionGreen}Junction Green's functions}

In this section we are going to calculate the junction Green's functions
that appeared in the calculation of the properties of the oscillator
stationary state, the average current and the noise. The junction
Green's functions encountered are \begin{eqnarray}
\Pi_{uu'}^{+}(\tau_{1},\tau_{2}) & = & -i\left\langle \mathrm{T}_{c}\left(h_{u}(\tau_{1})h_{u'}(\tau_{2})\right)\right\rangle ,\nonumber \\
\Pi_{uu'}^{-}(\tau_{1},\tau_{2}) & = & -\left\langle \mathrm{T}_{c}\left(h_{u}(\tau_{1})j_{u'}(\tau_{2})\right)\right\rangle ,\label{Pis}\end{eqnarray}
 where the notation was introduced in section \ref{definitions},
and the contour times discussed in section \ref{stationarystate}.
In terms of the tunneling operators $\mathcal{T}$ and $\mathcal{T}^{\dagger}$,
Eq.~(\ref{hamtum1}), taken in the interaction picture, we have \begin{eqnarray}
\Pi_{uu'}^{{\pm}}(\tau_{1},\tau_{2}) & = & -i\left[\left\langle \mathrm{T}_{c}\left(\mathcal{T}_{u}(\tau_{1})\mathcal{T}_{u'}^{\dagger}(\tau_{2})\right)\right\rangle \right.\nonumber \\
 & \pm & \left.\left\langle \mathrm{T}_{c}\left(\mathcal{T}_{u}^{\dagger}(\tau_{1})\mathcal{T}_{u'}(\tau_{2})\right)\right\rangle \right].\label{buvdef}\end{eqnarray}
 Inserting the tunneling operator from Eq\@.~(\ref{hu}) we get
the symmetric and antisymmetric combinations of left and right electrode
electron Green's functions \begin{eqnarray}
\Pi_{uu'}^{{\pm}}(\tau_{1},\tau_{2}) & = & -i\sum_{\mathbf{lr}}u_{\mathbf{lr}}u'_{\mathbf{rl}}\left[G_{\mathbf{l}}(\tau_{2}-\tau_{1})G_{\mathbf{r}}(\tau_{1}-\tau_{2})\right.\nonumber \\
 &  & \quad\quad\left.\pm G_{\mathbf{r}}(\tau_{2}-\tau_{1})G_{\mathbf{l}}(\tau_{1}-\tau_{2})\right],\label{Buv}\end{eqnarray}
 where we introduced the electron Green's function for the right and
left electrodes, for example \begin{equation}
G_{\mathbf{l}}(\tau_{1},\tau_{2})=-i\left\langle \mathrm{T}_{c}\left(c_{\mathbf{l}}(\tau_{1})c_{\mathbf{l}}^{\dagger}(\tau_{2})\right)\right\rangle .\end{equation}
 The Green's functions of interest can be obtained by standard techniques.
For example, \[
\Pi_{uu'\pm}^{<}(t_{1},t_{2})=-i\left[\left\langle \mathcal{T}_{u'}^{\dagger}(t_{2})\mathcal{T}_{u}(t_{1})\right\rangle \pm\left\langle \mathcal{T}_{u'}(t_{2})\mathcal{T}_{u}^{\dagger}(t_{1})\right\rangle \right],\]
 is given in terms of its Fourier transform \begin{eqnarray}
\Pi_{uu'\pm}^{<}(\omega) & = & -2\pi i\sum_{\mathbf{lr}}u_{\mathbf{lr}}u'_{\mathbf{rl}}\left[f(\epsilon_{\mathbf{l}})-f(\epsilon_{\mathbf{r}})\right]\nonumber \\
 &  & \times\left\{ \left[1+n(\epsilon_{\mathbf{r}}-\epsilon_{\mathbf{l}})\right]\delta(\epsilon_{\mathbf{r}}-\epsilon_{\mathbf{l}}+V-\omega)\right.\nonumber \\
 &  & \pm\left.n(\epsilon_{\mathbf{r}}-\epsilon_{\mathbf{l}})\delta(\epsilon_{\mathbf{r}}-\epsilon_{\mathbf{l}}+V+\omega)\right\} ,\end{eqnarray}
 where $n(\omega)$ is the Bose function. For voltages $V\ll E_{F}$
we obtain the approximate relation \begin{eqnarray}
 & \sum_{\mathbf{lr}}u_{\mathbf{lr}}u'_{\mathbf{rl}}\left[f(\epsilon_{\mathbf{l}})-f(\epsilon_{\mathbf{r}})\right]\delta(\epsilon_{\mathbf{r}}-\epsilon_{\mathbf{l}}+V)\nonumber \\
 & =V\,\sum_{\mathbf{lr}}u_{\mathbf{lr}}u'_{\mathbf{rl}}\left(-\frac{\partial f(\epsilon_{\mathbf{l}})}{\partial\epsilon_{\mathbf{l}}}\right)\delta(\epsilon_{\mathbf{r}}-\epsilon_{\mathbf{l}}),\end{eqnarray}
 as tunneling amplitudes depend only weakly on their arguments, so
that \begin{eqnarray*}
\Pi_{uu'\pm}^{<}(\omega) & = & iG_{uu'}\left[\left(V-\omega\right)\left[1+n(V-\omega)\right]\right.\\
 &  & \left.\pm\left(V+\omega\right)n(V+\omega)\right]\end{eqnarray*}
 where we introduced the conductance \begin{equation}
G_{uu'}=2\pi\sum_{\mathbf{lr}}u_{\mathbf{lr}}u'_{\mathbf{rl}}\left(-\frac{\partial f(\epsilon_{\mathbf{l}})}{\partial\epsilon_{\mathbf{l}}}\right)\delta(\epsilon_{\mathbf{r}}-\epsilon_{\mathbf{l}}).\end{equation}

The retarded, advanced and Keldysh bubbles \begin{eqnarray}
\Pi_{{uu'\pm}}^{R}(t_{1}-t_{2}) & = & -i\theta(t_{1}-t_{2})\left[\left\langle \left[\mathcal{T}_{u}(t_{1}),\mathcal{T}_{u'}^{\dagger}(t_{2})\right]\right\rangle \right.\nonumber \\
 &  & \left.\pm\left\langle \left[\mathcal{T}_{u}^{\dagger}(t_{1}),\mathcal{T}_{u'}(t_{2})\right]\right\rangle \right]\nonumber \\
\Pi_{{uu'\pm}}^{A}(t_{1}-t_{2}) & = & -i\theta(t_{2}-t_{1})\left[\left\langle \left[\mathcal{T}_{u}(t_{1}),\mathcal{T}_{u'}^{\dagger}(t_{2})\right]\right\rangle \right.\nonumber \\
 &  & \left.\pm\left\langle \left[\mathcal{T}_{u}^{\dagger}(t_{1}),\mathcal{T}_{u'}(t_{2})\right]\right\rangle \right]\nonumber \\
\Pi_{{uu'\pm}}^{K}(t_{1}-t_{2}) & = & -i\left[\left\langle \left\{ \mathcal{T}_{u}(t_{1}),\mathcal{T}_{u'}^{\dagger}(t_{2})\right\} \right\rangle \right.\nonumber \\
 &  & \left.\pm\left\langle \left\{ \mathcal{T}_{u}^{\dagger}(t_{1}),\mathcal{T}_{u'}(t_{2})\right\} \right\rangle \right]\label{PiRKA}\end{eqnarray}
 can be calculated in a similar way to give \begin{equation}
\Pi_{uu'}^{-}(\omega)=-i\left(\begin{array}{cc}
\Pi_{-}^{R}(\omega) & \Pi_{-}^{K}(\omega)\\
0 & \Pi_{-}^{A}(\omega)\end{array}\right)_{uu'}\end{equation}
 with \begin{eqnarray}
\Pi_{uu'-}^{R}(\omega) & = & -iG_{uu'}V+R_{uu'}^{-}(\omega)\nonumber \\
\Pi_{uu'-}^{K}(\omega) & = & -2iG_{uu'}\Delta_{V}(\omega)\label{PiMinus}\end{eqnarray}
 and the advanced part given by $\Pi_{-}^{A}(\omega)=\left[\Pi_{-}^{A}(\omega)\right]^{*}$,
with the functions $S_{V}(\omega)$ and $\Delta_{V}(\omega)$ defined
in Eqs.~(\ref{SV}, \ref{DeltaV}) respectively, and similarly $\Pi_{uu'}^{+}(\omega)$
is given by Eq.~(\ref{Greenb}).

The real parts of the retarded $\Pi_{uu'}^{\pm}(\omega)$ are given
by \begin{equation}
R_{uu'}^{\pm}(\omega)=2\pi\sum_{\mathbf{lr}}u_{\mathbf{lr}}u'_{\mathbf{rl}}\left[f(\epsilon_{\mathbf{l}})-f(\epsilon_{\mathbf{r}})\right]\left(P_{-}\pm P_{+}\right)\label{RB}\end{equation}
 with $P_{\pm}=1/(\epsilon_{\mathbf{l}}-\epsilon_{\mathbf{r}}\pm V+\omega).$
To estimate the real parts, we ignore the weak energy dependence of
the transmission amplitudes and the electron density of states are
constants up to a cut-off frequency of the order of the Fermi energy
$E_{F}$. For frequencies and voltages much smaller than the cutoff
frequency, $\omega,V\ll E_{F}$, we can approximate the reactive parts
of the response functions Eq.~(\ref{RB}) to give \begin{eqnarray}
R_{uu'}^{-}(\omega) & \sim & G_{uu'}V\frac{\omega}{E_{F}}\nonumber \\
R_{uu'}^{+}(\omega) & \sim & G_{uu'}E_{F}.\label{Rapprox}\end{eqnarray}
 This concludes our calculation of the junction Green's functions.

\section{Fourth order diagrams \label{fourthorderappendix}}

In this appendix we give the algebraic expressions for the fourth
order diagrams contributing to the current-current correlator. Let
us consider, for example, the contribution containing bubbles that
is linear in the oscillator Green's function, \textit{i.e.}, the diagrams
depicted in Fig.~\ref{vvwwpeak} \begin{eqnarray}
S_{v^{2}w^{2}}^{<}(\omega) & = & -i\sum_{kl}s^{kl}\left(\Pi_{-}^{l1}(\omega)\Pi_{-}^{k2}(\omega)D^{kl}(\omega)\right.\nonumber \\
 &  & +\Pi_{-}^{l1}(0)\Pi_{-}^{k2}(0)D^{12}(\omega)\nonumber \\
 &  & +\Pi_{-}^{l1}(\omega)\Pi_{-}^{k2}(0)D^{k2}(\omega)\nonumber \\
 &  & \left.+\Pi_{-}^{l1}(0)\Pi_{-}^{k2}(\omega)D^{k1}(\omega)\right),\label{StildevvwwofOmega}\end{eqnarray}
 where the full oscillator Green's function appears, accounting for
the influence of the interaction with the tunnel junction and the
matrix \begin{equation}
s^{kl}=\left(\begin{array}{cc}
1 & -1\\
-1 & 1\end{array}\right)\end{equation}
 introduces the correct signs for forward and backward contour (to
simplify notation, we suppress subscripts $v$ and $w$ on the right
hand side). Performing the summation over the contour indices $k$
and $l$, and inserting the bubbles and Green's functions we obtain
Eq.~(\ref{SvvwwRes}).

Similarly, the contribution from diagrams containing two oscillator
Green's functions, see Fig.~\ref{wwwwpeak}, becomes \begin{eqnarray}
S_{w^{4}}^{<}(\omega) & = & \sum_{kl}s^{kl}\int d\omega_{1}\Lambda^{kl}(\omega,\omega_{1}),\label{SwwwwApp}\\
\Lambda^{kl}(\omega,\omega_{1}) & = & \Pi_{-}^{l1}(\omega_{1})\Pi_{-}^{k2}(\omega_{1})D^{kl}(\omega_{1})D^{21}(\omega-\omega_{1})\nonumber \\
 & + & \Pi_{-}^{l1}(\omega-\omega_{1})\Pi_{-}^{k2}(\omega_{1})D^{2l}(\omega-\omega_{1})D^{k1}(\omega_{1}).\nonumber \end{eqnarray}
 Doing the summation and inserting the bubbles and oscillator Green's
functions we obtain Eq.~(\ref{SwwwwRes}), where we neglected a non-resonant
contribution that is small compared to the second order contributions
to the noise floor, Eqs.~(\ref{kwe}, \ref{Svvless}).

All diagrams not containing bubbles are of the non-resonant kind,
i.e. oscillator Green's functions always appear under frequency integration.
A typical non-bubble contribution to the noise spectrum is shown in
Fig.~\ref{nonbubblediagram} and given by\begin{widetext} \begin{equation}
\bar{s}_{v^{2}w^{2}}^{<}(\omega)=\int d\omega_{1}d\omega_{2}\sum_{kl}s^{kl}\sum_{\mathbf{l}_{1}\mathbf{r}_{1}}\sum_{\mathbf{l}_{2}\mathbf{r}_{2}}v^{2}w^{2}G_{\mathbf{l}_{1}}^{1k}(\omega_{1}-\omega_{2})G_{\mathbf{l}_{2}}^{2l}(\omega_{1}-\omega)G_{\mathbf{r}_{1}}^{k2}(\omega_{1})G_{\mathbf{r}_{2}}^{l1}(\omega_{1}-\omega)D^{k1}(\omega_{2}).\label{sbar}\end{equation}
 \end{widetext} Since there are no resonant peaks present, contributions
of this type have to be compared to the second order contribution
to the noise floor, Eqs.~(\ref{Svvless}, \ref{kwe}) and since they
are always of higher order in tunneling they can be neglected.


\begin{thebibliography}{10}
\bibitem{SchwRou05}K. C. Schwab and M. L. Roukes, Physics Today,
July 2005, 37 (2005).

\bibitem{Ble04}M. Blencowe, Phys. Rep. \textbf{395}, 159 (2004).

\bibitem{EkiRou05}K. L. Ekinci and M. L. Roukes, Review of Scientific
Instruments \textbf{76}, 061101 (2005).

\bibitem{Saz04}V. Sazonova, Y. Yaish, H. Üstünel, D. Roundy, T. A.
Arias and P. L. McEuen, Nature \textbf{431}, 284 (2004).

\bibitem{Leg02}A. J. Leggett, J. Phys.: Condens. Matter \textbf{14},
R415-R451 (2002).

\bibitem{MarSimPen03}W. Marshall, C. Simon, R. Penrose and D. Bouwmeester,
Phys. Rev. Lett. \textbf{91}, 130401 (2003).

\bibitem{NaiBuuLah06}A. Naik, O. Buu, M. D. LaHaye, A. D. Armour,
A. A. Clerk, M. P. Blencowe and K. C. Schwab, Nature \textbf{443},
193 (2006).

\bibitem{LahBuuCam04}M. D. LaHaye, O. Buu, B. Camarota and K. C.
Schwab, Science \textbf{304}, 74 (2004).

\bibitem{GigBohZei06}S. Gigan, H. R. Böhm, M. Paternostro, F. Blaser,
G. Langer, J. B. Hertzberg, K. C. Schwab, D. Bäuerle, M. Aspelmeyer
and A. Zeilinger, Nature \textbf{444}, 67 (2006).

\bibitem{ArcCohBri06}O. Arcizet, P.-F. Cohandon, T. Briant, M. Pinard,
and A. Heidmann, Nature \textbf{444}, 71 (2006).

\bibitem{KleBou06}D. Kleckner and D. Bouwmeester, Nature \textbf{444},
75 (2006).

\bibitem{YanCalFen06}Y. T. Yang, C. Callegari, X. L. Feng, K. L.
Ekinci and M. L. Roukes, Nano Letters \textbf{6}, 583 (2006).

\bibitem{RugBudMam04}D. Rugar, R. Budakian, H. J. Mamin and B. W.
Chu, Nature \textbf{430}, 329-332 (2004).

\bibitem{GelCle05}M. R. Geller and A. N. Cleland, Phys. Rev. A \textbf{71},
032311 (2005).

\bibitem{MozMar02}D. Mozyrsky and I. Martin, Phys. Rev. Lett. \textbf{89},
018301 (2002).

\bibitem{CleGir04}A. A. Clerk and S. M. Girvin, Phys. Rev. \textbf{70},
121303 (2004).

\bibitem{WabKhoRam05}J. Wabnig, D. V. Khomitsky, J. Rammer and A.
L. Shelankov, Phys. Rev. B \textbf{72}, 165347 (2005).

\bibitem{SmiMouHor03}A. Yu. Smirnov, L. G. Mourokh and N. J. M. Horing,
Phys. Rev. B \textbf{67}, 115312 (2003).

\bibitem{RodArm05}D. A. Rodrigues and A. D. Armour, Phys. Rev. B
\textbf{72}, 085324 (2005).

\bibitem{BleImbArm05}M. P. Blencowe, J. Imbers and A. D. Armour,
New Journal of Physics \textbf{7}, 236 (2005).

\bibitem{CleBen05}A. A. Clerk and S. Bennett, New Journal of Physics
\textbf{7}, 238 (2005).

\bibitem{SpiLehSid03}L. Spietz, K. W. Lehnert, I. Siddiqi and R.
J. Schoelkopf, Science \textbf{300}, 1929 (2003).

\bibitem{LiCuiYan05}Xin-Qi Li, Ping Cui and Yi Jing Yan, Phys. Rev.
Lett. \textbf{94}, 066803 (2005).

\bibitem{LesLoo97}G. B. Lesovik and R. Loosen, JETP Lett. \textbf{65},
295 (1997).

\bibitem{AguKou00}R. Aguado and L. P. Kouwenhoven, Phys. Rev. Lett.
\textbf{84}, 1986 (2000).

\bibitem{GavLevImr00}U. Gavish, Y. Levinson and Y. Imry, Phys. Rev.
B \textbf{62}, R10637 (2000).

\bibitem{RamSmi86}J. Rammer and H. Smith, Rev. Mod. Phys. \textbf{58},
323 (1986).

\bibitem{Kog96}Sh. Kogan, \emph{Electronic noise and fluctuations
in solids}, Cambridge University Press, (1996).

\bibitem{Cle04}A. A. Clerk, Phys. Rev. B \textbf{70}, 245306 (2004).

\bibitem{DahDenLan69}A. J. Dahm, A. Denenstein, D. N. Langenberg,
W. H. Parker, D. Rogovin and D. J. Scalapino, Phys. Rev. Lett. \textbf{22},
1416 (1969).

\bibitem{bum} The additional pumping contribution to the current
found in Ref.~\onlinecite{WabKhoRam05} is not present here, since
for simplicity we take the junction to be symmetric implying real
transmission matrix elements. 
\end{thebibliography}
\end{document}